\documentclass[twocolumn, prd, superscriptaddress,nofootinbib]{revtex4-1}

\usepackage{array,mathtools,booktabs}
\newcolumntype{C}{>{$}c<{$}}
\AtBeginDocument{
\heavyrulewidth=.08em
\lightrulewidth=.05em
\cmidrulewidth=.03em
\belowrulesep=.65ex
\belowbottomsep=0pt
\aboverulesep=.4ex
\abovetopsep=0pt
\cmidrulesep=\doublerulesep
\cmidrulekern=.5em
\defaultaddspace=.5em
}

\usepackage{appendix}
\usepackage[utf8]{inputenc}
\usepackage[dvipsnames]{xcolor}
\usepackage{amsmath,amssymb}
\usepackage{hyperref}
\hypersetup{
    colorlinks,
    linkcolor={red!80!black},
    citecolor={green!60!black},
    urlcolor={blue!60!black}
}
\usepackage{graphicx}

\newcommand{\Eboom}{\mathcal{E}_\x{boom}}
\newcommand{\OO}{\mathcal{O}}

\newcommand{\TeV}{\text{TeV}}
\newcommand{\GeV}{\text{GeV}}
\newcommand{\MeV}{\text{MeV}}
\newcommand{\keV}{\text{keV}}
\newcommand{\cm}{\text{cm}}

\newcommand{\Mpl}{M_{\text{pl}}}

\newcommand{\x}[1]{\ensuremath{\text{#1}}} % text mode shortcut

\newcommand{\ion}{\text{ion}}

\def\r{\right)}
\def\l{\left(}

\newcounter{qnumber}

\newcounter{qnumber2}

\newcounter{qnumber3}

\definecolor{c1}{rgb}{0.368417, 0.506779, 0.709798}
\definecolor{c2}{rgb}{0.880722, 0.611041, 0.142051}
\definecolor{c3}{rgb}{0.560181, 0.691569, 0.194885}
\definecolor{c4}{rgb}{0.922526, 0.385626, 0.209179}
\definecolor{c5}{rgb}{0.528488, 0.470624, 0.701351}
\definecolor{c6}{rgb}{0.772079, 0.431554, 0.102387}
\definecolor{c7}{rgb}{0.363898, 0.618501, 0.782349}
\definecolor{c8}{rgb}{1., 0.75, 0.}
\definecolor{c9}{rgb}{0.647624, 0.37816, 0.614037}
\definecolor{c10}{rgb}{0.571589, 0.586483, 0.}
\definecolor{c11}{rgb}{0.915, 0.3325, 0.2125}
\definecolor{c12}{rgb}{0.400822, 0.522007, 0.85}
\definecolor{c13}{rgb}{0.972829, 0.621644, 0.073362}
\definecolor{c14}{rgb}{0.736783, 0.358, 0.503027}
\definecolor{c15}{rgb}{0.280264, 0.715, 0.429209}

\begin{document}

%\preprint{APS/123-QED}

\title{Type Ia supernovae from dark matter core collapse}

\author{Ryan Janish}
\affiliation{Berkeley Center for Theoretical Physics, Department of Physics,
University of California, Berkeley, CA 94720, USA}

\author{Vijay Narayan}
\affiliation{Berkeley Center for Theoretical Physics, Department of Physics,
University of California, Berkeley, CA 94720, USA}

\author{Paul Riggins}
\affiliation{Berkeley Center for Theoretical Physics, Department of Physics,
University of California, Berkeley, CA 94720, USA}

\begin{abstract}

Dark matter (DM) which sufficiently heats a local region in a white dwarf will trigger runaway fusion, igniting a type Ia supernova (SN).
In a companion paper, this instability was used to constrain DM heavier than $10^{16}$ GeV which ignites SN through the violent interaction of one or two individual DM particles with the stellar medium.
Here we study the ignition of supernovae by the formation and self-gravitational collapse of a DM core containing many DM particles.
For non-annihilating DM, such a core collapse may lead to a mini black hole that can ignite SN through the emission of Hawking radiation, or possibly as a by-product of accretion.
For annihilating DM, core collapse leads to an increasing annihilation rate and can ignite SN through a large number of rapid annihilations.
These processes extend the previously derived constraints on DM to masses as low as $10^{5}$ GeV.

\end{abstract}

\maketitle

\section{Introduction}
\label{sec:intro}

Dark matter (DM) accounts for over 80\% of the matter density of the Universe, but its identity remains unknown.
While direct detection~\cite{Aprile:2018dbl} is a promising approach to identifying the nature of DM, searches for indirect signatures of DM interactions in astrophysical systems is also fruitful, particularly if the unknown DM mass happens to be large.

It was recently suggested~\cite{Graham:2015apa} that white dwarfs (WD) act as astrophysical DM detectors: DM may heat a local region of a WD and trigger thermonuclear runaway fusion, resulting in a type Ia supernova (SN).
DM ignition of sub-Chandrasekhar WDs was further studied in a companion paper~\cite{Graham:2018efk}, where we showed that generic classes of DM capable of producing high-energy standard model (SM) particles in the star can be constrained, e.g., by DM annihilations or decay to SM products.
As an illustrative example,~\cite{Graham:2018efk} placed new constraints on ultra-heavy DM with masses greater than $10^{16}~\GeV$ for which a single annihilation or decay is sufficient to ignite a SN.

Here we examine the possibility of igniting SN by the formation and self-gravitational collapse of a DM core.
We study two novel processes by which a collapsing DM core in a WD can ignite a SN---these were first pointed out in~\cite{Graham:2018efk}, and are studied here in more detail.
If the DM has negligible annihilation cross section, so-called asymmetric DM, collapse may result in a mini black hole (BH) that can ignite a SN via the emission of energetic Hawking radiation or possibly as it accretes.
If the DM has a small but non-zero annihilation cross section, collapse can dramatically increase the number density of the DM core, resulting in SN ignition via a large number of rapid annihilations.
Both of these processes extend the previously derived constraints on DM in~\cite{Graham:2018efk}, notably to masses as low as $10^{5}~\GeV$.

A number of potential observables of DM cores in compact objects have been considered in the literature.
These include:
(1) gravitational effects of DM cores on the structure of low-mass stars~\cite{Bottino:2002pd, Cumberbatch:2010hh, PhysRevLett.105.011301, Lopes:2012af, Casanellas:2012jp}, WDs~\cite{Leung:2013pra}, and neutron stars (NS)~\cite{Sandin:2008db, Ciarcelluti:2010ji, Li:2012ii, Goldman:2013qla},
(2) BH formation and subsequent destruction of host NSs~\cite{PhysRevD.40.3221, Gould:1989gw, PhysRevD.83.083512, PhysRevLett.107.091301, McDermott:2011jp, Jamison:2013yya, PhysRevD.87.123537, Bramante:2013hn, Kouvaris:2013kra, PhysRevD.87.123507, PhysRevLett.108.191301, Bramante:2013nma, Bramante:2014zca, Bramante:2017ulk}, and
(3) anomalous heating from DM annihilations or scatters in WDs and NSs~\cite{PhysRevD.81.083520, PhysRevD.81.103531, PhysRevD.77.023006, PhysRevD.81.123521, PhysRevD.82.063531,PhysRevD.77.043515, Baryakhtar:2017dbj, PhysRevD.97.043006}.
See also~\cite{PhysRevLett.115.111301, Brito:2015yfh} for unique astroseismology signatures of possible low-mass bosonic DM cores.
We emphasize that the signature of a DM core igniting a type Ia SN is distinct from these, and thus the constraints derived here are complementary.
For instance, while it has been known that DM cores which form evaporating mini BHs are practically unobservable in a NS, this is decidedly not the case in a WD where (as we show) such BHs will typically ignite a SN.
Note that~\cite{Bramante:2015cua} considers DM cores in WDs which inject heat and ignite SN through elastic DM-nuclear scatters---we discuss this process in more detail later as it pertains to our new constraints.

The paper is organized as follows.
In Sec.~\ref{sec:startSN}, we review the triggering of runaway fusion by localized energy deposition in a WD.
In Sec.~\ref{sec:corecollapse}, we summarize the necessary conditions for DM core collapse and discuss the generic end-states of such collapse.
In Sec.~\ref{sec:BH} and Sec.~\ref{sec:annburst}, we derive constraints on DM cores which would ignite SN by the processes described above, namely black hole formation and DM-DM annihilations. We conclude in Sec.~\ref{sec:discussion}.

\section{Triggering thermonuclear runaway}
\label{sec:startSN}

Thermonuclear runaway in a carbon WD generally occurs when the cooling timescale of a hot region exceeds the fusion timescale.
Cooling is dominated by the thermal diffusion of either photons or degenerate electrons, while the highly exothermic fusion of carbon ions is unsuppressed at temperatures greater than their Coulomb threshold $\approx \MeV$.
Crucially, the diffusion time increases with the size $L$ of the heated region while the fusion time is independent of $L$.
This defines a critical trigger size and temperature for ignition:
\begin{equation}
\label{eq:boom1}
L \gtrsim \lambda_T ~~\x{and}~~T \gtrsim \MeV ~ \Rightarrow \x{ignite supernova}.
\end{equation}
$\lambda_T$ was numerically computed in~\cite{woosley} and is $\lambda_T \approx 10^{-5}~\cm$ at a number density $n_\x{ion} \approx 10^{32}~\cm^{-3}$.

One can also consider, as in~\cite{Graham:2018efk}, the critical energy $\Eboom$ required to heat an entire trigger region $\lambda_T^3$ to an $\MeV$.
$\Eboom \approx 10^{16}~\GeV$ for $n_\x{ion} \approx 10^{32}~\cm^{-3}$ and sharply increases at lower WD densities---this agrees with the expectation that WDs grow closer to instability as they approach the Chandrasekhar mass.
Of course to trigger runaway fusion, an energy in excess of $\Eboom$ must also be deposited sufficiently rapidly.
The relevant timescale is the characteristic diffusion time $\tau_\x{diff}$ across a region of size $\lambda_T$ at a temperature $\approx \MeV$.
This diffusion time is also computed in~\cite{woosley} to be $\tau_\x{diff} \approx 10^{-12}~\x{s}$ at densities $n_\x{ion} \approx 10^{32}~\cm^{-3}$.
Therefore a total energy $\mathcal{E}$, specifically deposited within a trigger region $\lesssim \lambda_T^3$ and a diffusion time $\lesssim \tau_\x{diff}$, will ignite a SN if:
\begin{equation}
\label{eq:boom2}
\mathcal{E} \gtrsim \Eboom ~ \Rightarrow \x{ignite supernova}.
\end{equation}

One possibility is that the necessary energy~\eqref{eq:boom2} is deposited directly to carbon ions, e.g., by a transiting primordial BH~\cite{Graham:2015apa}.
It is also possible to deposit this energy indirectly, e.g., by DM interactions releasing SM particles into the stellar medium~\cite{Graham:2018efk}.
To this end the stopping distances of high-energy ($\gg \MeV$) particles in a WD was calculated in~\cite{Graham:2018efk}, where it was shown that hadrons, photons and electrons all transfer their energies to the stellar medium within a distance of order $\lambda_T$ (the sole exception being neutrinos).
We thus safely presume that any $\mathcal{E} \gtrsim \Eboom$ released into these SM products inside $\lambda_T^3$ will be efficiently deposited and thermalized within this region as well.

In summary, the rate of any process which deposits an energy $\mathcal{E}$ (defined to be localized spatially within $\lambda_T^3$ and temporally within $\tau_\x{diff}$) that satisfies~\eqref{eq:boom2} can be constrained.
This is done by either demanding that a single explosive event not occur during the lifetime of an observed heavy $\gtrsim 1.2~M_{\odot}$ WD%
\footnote{For instance, the Sloan Digital Sky survey has cataloged $> 10$ such heavy WDs~\cite{Kepler:2006ns}.},
or that the occurrence of many such events throughout the galaxy in predominantly lower mass WDs not affect the observed SN rate.
For simplicity we just utilize the former here and the existence of a WD with properties:
\begin{align}
\label{eq:wdparam}
&n_\x{ion} \approx 10^{31}~\cm^{-3}, ~~~ \rho_\x{WD} \approx  3 \cdot 10^{8}~\frac{\x{g}}{\cm^3} \nonumber, \\
&M_\x{WD} \approx 1.25~M_{\odot}, ~~~ R_\x{WD} \approx 3000~\x{km}.
\end{align}
Here $n_\x{ion}$ and $\rho_\x{WD}$ refer to the central density of the WD, and we relate this to its mass and radius using the equation of state formulated in~\cite{cococubed}.
While the average density is smaller by a factor $\sim 10^{-1}$, $n_\x{ion}$ only changes by $\OO(1)$ from the central value out to distances $\sim R_\x{WD}/2$~\cite{chandra}.
For such a WD, the relevant trigger scales are of order:
\begin{align}
\label{eq:wdparam2}
&\lambda_T \approx 4 \cdot 10^{-5}~\cm, \nonumber \\
&\Eboom \approx 7 \cdot 10^{16}~\GeV, \nonumber \\
&\tau_\x{diff} \approx 4 \cdot 10^{-11}~\x{s}
\end{align}
These values are approximate, but we expect they are accurate at the order of magnitude level, as are the ensuing constraints.
Finally, we assume the WD has a typical interior temperature $T_\x{WD} \approx \keV$ and lifetime $\tau_\x{WD} \approx 5~\x{Gyr}$ \cite{kippenhan}.%
\footnote{The age of a WD is typically estimated by measuring its temperature and modeling the cooling over time.}
%Throughout we will use natural units: $c = \hbar = k_B = 1$.

\section{Dark matter core collapse}
\label{sec:corecollapse}

Here we review the conditions for DM capture, collection, and self-gravitational collapse in a WD.
As much of this discussion is already present in the literature, in what follows we simply quote the relevant results.
We assume throughout that the DM loses energy primarily by short-range nuclear scatters.
While other dissipation mechanisms are certainly possible (such as exciting dark states or emitting radiation) we will not treat these here.

Consider DM with mass $m_\chi \gg 10~\GeV$ and scattering cross section off ions $\sigma_{\chi A}$.
For spin-independent interactions, $\sigma_{\chi A}$ is related to the DM-nucleon cross section $\sigma_{\chi n}$ by
\begin{equation}
\label{eq:Helm}
\sigma_{\chi A} = A^2 \l \frac{\mu_{\chi A}}{\mu_{\chi n}}\r^2 F^2(q) \sigma_{\chi n} \sim A^4 F^2(q) \sigma_{\chi n},
\end{equation}
where $F^2(q)$ is the Helm form factor~\cite{PhysRev.104.1466}, and $q \sim m_\text{ion} v_\x{rel} \sim m_\x{ion} \x{max}[v, v_\x{ion}]$ is the momentum transfer between the DM at velocity $v$ and a nuclear target.
Currently the most stringent constraints on $\sigma_{\chi n}$ come from Xenon 1T~\cite{Aprile:2018dbl}:
\begin{equation}
\label{eq:xenon1T}
\sigma_{\chi n} < 10^{-45}~\cm^2 \l \frac{m_\chi}{\TeV} \r,
\end{equation}
It is also possible for DM to have spin-dependent interactions (e.g., Majorana DM) which does not benefit from a $A^2$ coherent enhancement and is less constrained by direct detection~\cite{Aprile:2019dbj}.
WDs predominantly consist of spin-zero nuclei (${}^{12}$C, ${}^{16}$O), though as pointed out by~\cite{PhysRevD.83.083512} DM capture/thermalization can proceed by scattering off a lower density of non-zero spin nuclei (e.g., a small fraction of ${}^{13}$C).
For simplicity, we will restrict our attention here only to spin-independent interactions.

\subsection{Core formation}

DM capture in compact objects has a long history~\cite{Press:1985ug, Gould:1987ir}, though the usual formulae must be modified to account for heavy DM requiring multiple scatters to be captured (e.g., see~\cite{Graham:2018efk}).
DM transits the WD at a rate
\begin{align}
\label{eq:transitrate}
  \Gamma_\x{trans} \sim
  \frac{\rho_{\chi}}{m_\chi} R_\x{WD}^2
  \l\frac{v_\x{esc}}{v_\x{halo}}\r^2 v_\x{halo},
\end{align}
where $v_\x{esc} \approx 2 \cdot 10^{-2}$ is the escape velocity and $v_\x{halo} \approx 10^{-3}$ is the virial velocity of our galactic halo.
$\rho_\chi$ is the DM density in the region of the WD---we may consider either nearby WDs~\cite{Kepler:2006ns} with $\rho_\chi \approx 0.4~\frac{\GeV}{\cm^3}$ or WDs close to the galactic center~\cite{Perez:2015aa} where it is expected that $\rho_\chi \gtrsim 10^3~\frac{\GeV}{\cm^3}$~\cite{Nesti:2013uwa}.
Meanwhile, DM is captured by the WD at rate that is parametrically
\begin{align}
\label{eq:capturerate}
  \Gamma_\x{cap} \sim \Gamma_\x{trans} \cdot
  \x{min}\left [1,~\frac{N_\x{scat}}{N_\x{cap}(v_\x{halo})} \right ].
\end{align}
$N_\x{scat} \sim n_\x{ion} \sigma_{\chi A} R_\x{WD}$ is the average number of DM scatters during a single transit, and $N_\x{cap}(v) \sim \frac{m_\chi v^2}{m_\x{ion} v_\x{esc}^2}$ is roughly the number of scatters needed for DM with velocity $v$ asymptotically far away from star to become gravitationally bound, though with a necessary minimum of $N_\x{cap} \geq 1$.
More properly, $\Gamma_\x{cap}$ should be numerically calculated~\cite{Bramante:2017xlb}, though the expression in~\eqref{eq:capturerate} is parametrically correct.
Based on the assumed WD parameters~\eqref{eq:wdparam}, we find $N_\x{cap}(v_\x{halo}) > 1$ for DM masses $m_\chi > 10~\x{TeV}$; in this regime, the capture rate scales as $\Gamma_\x{cap} \propto \frac{\sigma_{\chi A}}{m_\chi^2}$ as opposed to the usual $\Gamma_\x{cap} \propto \frac{\sigma_{\chi A}}{m_\chi}$ result that is often used.

We now turn to DM thermalization.
For sufficiently small $\sigma_{\chi A}$, captured DM will follow gravitational orbits which gradually shrink as the DM dissipates energy and eventually reaches velocities $v_\text{th}$ and settles at a radius $R_\text{th}$ where its kinetic energy balances the gravitational potential of the enclosed WD mass:
\begin{align}
  \label{eq:Rth}
  v_\x{th} &\sim \sqrt{\frac{T_\text{WD}}{m_\chi}} \approx 10^{-7} \l \frac{m_\chi}{10^{8} ~\GeV}\r^{-1/2}, \\
  R_\x{th} &\sim \l \frac{T_\x{WD}}{G m_\chi \rho_\x{WD}}\r^{1/2}
 \approx 500~\cm \l \frac{m_\chi}{10^{8} ~\GeV}\r^{-1/2}.
\end{align}
Considering dissipation due to elastic nuclear scattering, the evolution to $R_\x{th}$ is predominately orbital if the gravitational dynamical timescale $t_\x{ff} \sim \left(G \rho_\text{WD}\right)^{-1/2} \approx 0.1~\x{s}$ is shorter than the time $t_\theta$ over which a DM particle is deflected from its orbital trajectory by ion scatters.
For $m_\chi \gg m_\x{ion}$, this deflection is the result of a Brownian process whereby many small momentum transfers $\delta p \sim m_\ion \; \x{max} [ v, v_\ion ]$ due to individual ion-DM scatters add incoherently to produce a net $\OO\l1\r$ deflection.
This requires a number of scatters
\begin{align}
    N_\theta \sim \l \frac{m_\chi v}{m_\ion \; \x{max} [ v, v_\ion ]} \r^2,
\end{align}
and occurs over a time $t_\theta \sim \frac{N_\theta}{n_\ion \sigma_{\chi A} \; \x{max} [ v, v_\ion ]}$.
We find $t_\theta$ exceeds $t_\x{ff}$ at a critical cross section
\begin{align}
\label{eq:sigma_theta}
 \sigma_\theta \sim \frac{m_\chi}{\rho_\x{WD} v_\ion  t_\x{ff}} \sim 10^{-39}~\cm^2 \l \frac{m_\chi}{\TeV} \r.
\end{align}

For $\sigma_{\chi A} < \sigma_\theta$, the thermalization time depends on the rate at which DM orbits dissipate their energy and decay.
For elastic nuclear scatters,
\begin{align}
\label{eq:dEdt}
    \frac{dE}{dt}  \sim \rho_\x{WD} \sigma_{\chi A}  v^2
    \; \x{max} [v, v_\ion ].
\end{align}
where $v_\ion \sim \sqrt{\frac{T_\x{WD}}{m_\x{ion}}} \approx 4 \cdot 10^{-4}$ is the thermal ion velocity and $v$ is the velocity of the ``in-falling" DM.
It is important to distinguish the rate of energy loss in the regimes of inertial and viscous drag, with the latter being relevant once $v$ drops below $v_\x{ion}$.

Thermalization proceeds in three stages (e.g., see~\cite{PhysRevD.83.083512} for a detailed derivation).
First, the DM may pass through the star many times on a wide elliptic orbit of initial size $R_i \gg R_\x{WD}$ set by the number of scatters during the first stellar transit:
\begin{align}
\label{eq:Ri}
  R_i \sim R_\x{WD} \l \frac{m_\chi}{m_\x{ion}} \r \frac{1}{\x{max} [N_\x{scat}, 1 ]}.
\end{align}
The time for the DM orbital size to become contained within the WD is then:
\begin{align}
\label{eq:t1}
  t_1 &\sim \frac{m_\chi}{\rho_\x{WD} \sigma_{\chi A} v_\x{esc}} \l \frac{R_i}{R_\x{WD}} \r^{1/2}, ~~~ (R_i \to R_\text{WD}).
\end{align}
Subsequently the DM completes many orbits within the star, losing energy according to~\eqref{eq:dEdt}.
The DM first slows to $v_\text{ion}$ in a time
\begin{align}
\label{eq:t2}
  t_2  \sim \frac{m_\chi}{\rho_\x{WD} \sigma_{\chi A} v_\x{ion}}, ~~~ (v_\text{esc} \to v_\x{ion}).
\end{align}
and then thermalizes at $v_\text{th}$ in a time that is logarithmically greater:
\begin{equation}
\label{eq:t3}
t_3 \sim t_2 \log \l \frac{m_\chi}{m_\ion} \r, ~~~ (v_\x{ion} \to v_\x{th}).
\end{equation}

The process of thermalization is qualitatively different if $\sigma_{\chi A} > \sigma_\theta$.
In this scenario, it firstly happens that $\OO(1)$ of the DM kinetic energy is lost in the first pass through the star.
Subsequently, elastic deflection will become important before a DM particle reaches $R_\x{th}$, and its motion will thus be Brownian with an inward gravitational drift.
%, described by a Langevin equation.
Since the DM now scatters frequently with ions, it becomes thermal even outside of $R_\text{th}$ and equilibrates with the stellar medium at temperature $T_\x{WD}$.
The DM then settles into a Boltzmann distribution, in this case a Gaussian density profile of size $R_\x{th}$ in the center of the star.
The timescale $t_\x{drift}$ to develop this profile is set by the rate at which thermal DM drifts inward to $R_\x{th}$, which is given by the local terminal speed of free-falling DM due to the collisional drag force $F_A$,
\begin{align}
    F_A \sim \rho_\x{WD} \sigma_{\chi A}  v \; \x{max} [v, v_\ion ].
\end{align}
Conservatively taking deflection to be relevant to the very edge of the star, we find
\begin{align}
    t_\x{drift} \sim t_\x{ff} \; \frac{\sigma_{\chi A}}{\sigma_\theta}
    \; \log\l\frac{R_\x{WD}}{R_\x{th}}\r.
\end{align}

In summary, for DM with $\sigma_{\chi A} < \sigma_\theta$, a core will form in a WD if
\begin{equation}
\label{eq:core}
t_1 + t_2 + t_3 < \tau_\x{WD} ~~~ \x{(form DM core)},
\end{equation}
whereas for $\sigma_{\chi A} > \sigma_\theta$, a core forms if
\begin{equation}
\label{eq:core-bigsigma}
t_\x{drift} < \tau_\x{WD} ~~~~~~~~~~~ \x{(form DM core)}.
\end{equation}
Note that the first condition~\eqref{eq:core} sets a lower threshold on $\sigma_{\chi A}$ at a given $m_\chi$ for the formation of a core, while the second condition~\eqref{eq:core-bigsigma} sets an upper bound roughly $\sigma_{\chi A} \lesssim 10^{16}~\sigma_\theta$.
This gives the full range of elastic cross-sections over which DM cores form.

\subsection{Asymmetric DM Collapse}
\label{sec:AsymmetricDM}

First consider the evolution of a core of non-annihilating DM, herein referred to as asymmetric DM~\cite{Nussinov:1985xr, Zurek:2013wia}.
Upon formation, the DM core will steadily collect at $R_\text{th}$ at a rate $\Gamma_\text{cap}$.
If its density ever exceeds the WD density $\rho_\x{WD}$, then the core will become self-gravitating.
The critical number of DM particles needed for the onset of self-gravitation is
\begin{align}
\label{eq:Nsg}
    N_\x{sg} \sim \frac{\rho_\x{WD} R^3_\x{th}}{m_\chi} \approx 5 \cdot 10^{32} \l \frac{m_\chi}{10^{8} ~\GeV} \r^{-5/2},
\end{align}
while the total number of DM particles that can possibly be collected within $\tau_\x{WD}$ is simply:
\begin{align}
\label{eq:Nlife}
    N_\x{life} \sim \Gamma_\x{cap} \tau_\x{WD}.
\end{align}
Thus self-gravitational collapse requires
\begin{equation}
\label{eq:sgcond}
N_\x{sg} < N_\x{life},~~~ \x{(core self-gravitates)}.
\end{equation}
This sets an upper limit on the DM mass that can form a self-gravitating core $m_\chi \gtrsim 100~\TeV$ (or $R_\x{th} \lesssim 0.1~\x{km}$), taking the maximum possible capture rate $\Gamma_\x{cap} = \Gamma_\x{trans}$ and $\rho_\chi = 0.4~\frac{\GeV}{\cm^3}$.
%At these masses, evaporation of DM particles on the tail of the Boltzmann distribution is clearly negligible.

Of course, this assumes that the DM core obeys Maxwell-Boltzmann statistics throughout the collection phase.
In general, the quantum statistics of DM with velocity $v$ in a core of size $N$ becomes important once the de Broglie wavelength of individual DM particles $\sim \frac{1}{m_\chi v}$ exceeds their physical separation in the core $\sim \frac{r}{N^{1/3}}$.
For the thermal DM population at $R_\x{th}$, this occurs after it has collected a number:
\begin{equation}
\label{eq:Nqm}
N_\x{QM, th} \sim (m_\chi T_\x{WD})^{3/2} R_\x{th}^3 \sim \frac{T_\x{WD}^3}{(G \rho_\x{WD})^{3/2}} \approx 10^{52},
\end{equation}
which is greater than $N_\x{sg}$ for all DM masses $m_\chi \gtrsim \GeV$.
In the case of bosonic DM, if the core reaches $N_\x{QM, th}$ before the onset of self-gravitation it will begin populating a Bose-Einstein condensate (BEC).
A more compact BEC could then self-gravitate earlier, as considered by~\cite{PhysRevD.40.3221, McDermott:2011jp, PhysRevLett.107.091301} in a NS.
We find this is not possible in a WD, namely $N_\x{QM, th} \gg N_\x{life}$ even for light bosonic DM $m_\chi \lesssim \GeV$.
Thus the condition for core collapse is indeed~\eqref{eq:sgcond}.

For simplicity, we focus on DM which scatters infrequently with the medium, $\sigma_{\chi A} < \sigma_\theta$, see~\eqref{eq:sigma_theta}.
The gravitational dynamical timescale of a collapsing core is always decreasing, so orbital motion would dominate over Brownian at some point during the collapse.
In the opposite regime a collapse will still occur, however its evolution qualitatively differs as it begins with an initial phase of inward Brownian motion in which the DM velocities may be parametrically smaller than in the orbital case.

\begin{figure}
\centering
\includegraphics[width=8cm]{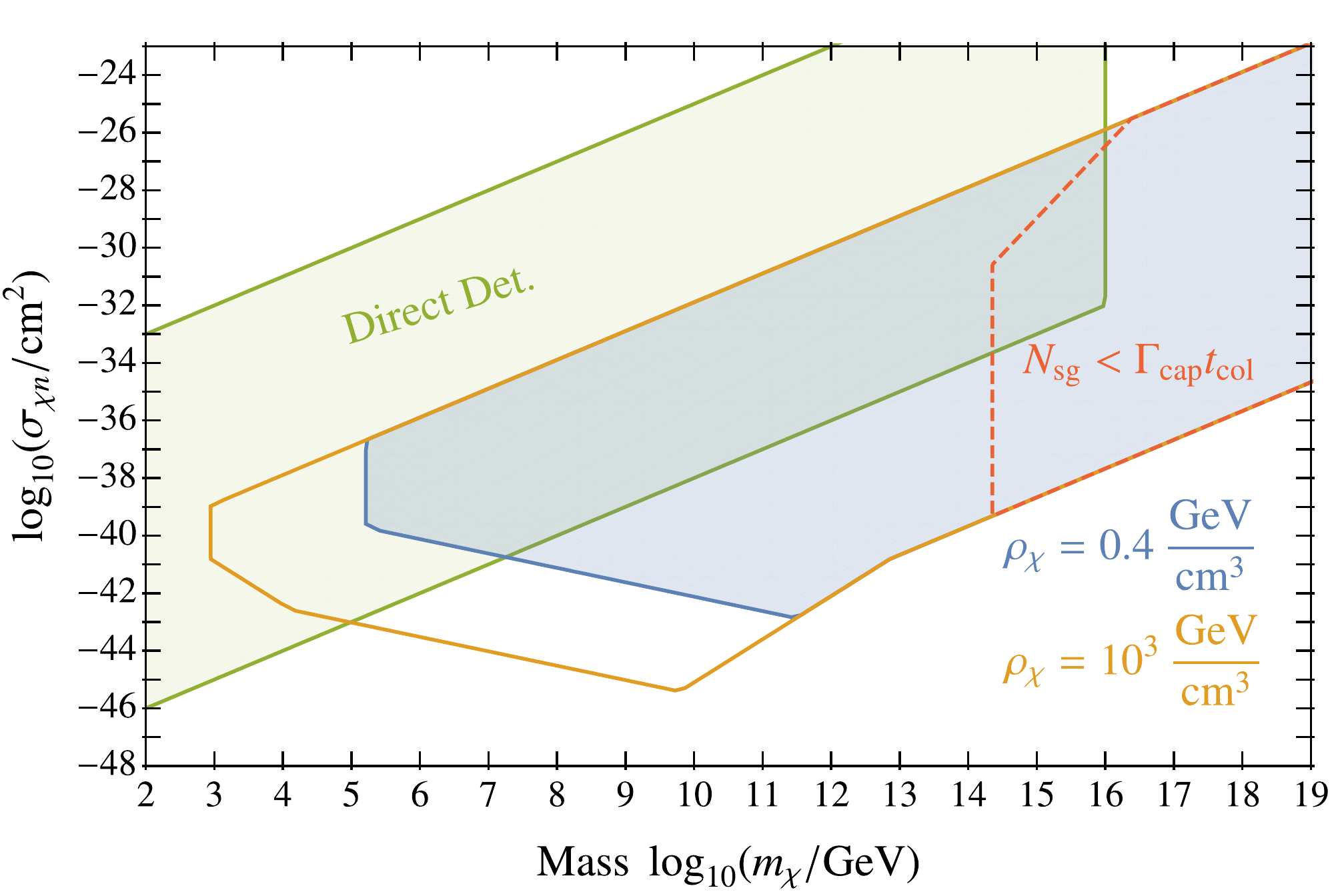}
\caption{Parameter space $\{ m_\chi, \sigma_{\chi n}\}$ of asymmetric DM in which a DM core forms and collapses within $\tau_\x{WD} \approx 5~\x{Gyr}$ in a WD of local DM density $\rho_\chi$.
See text for details.
}
\label{fig:collapsecond}
\end{figure}

In summary, the conditions~\eqref{eq:core} and~\eqref{eq:sgcond} on $\{m_\chi ,\sigma_{\chi n} \}$ parameter space for which a DM core forms and collapses in a WD are depicted in Fig.~\ref{fig:collapsecond}.
We also show a rough amalgamation (e.g., see~\cite{Mack:2007xj}), extending to large DM masses and cross sections, of the constraints from underground direct detection experiments including Xenon 1T~\cite{Aprile:2018dbl}.

We now turn to the dynamics of collapse.
In order for a self-gravitating DM core to shrink, it must lose the excess gravitational potential energy.
The ``cooling" timescale $t_\x{col}$ (leading to gravitational heating of the DM) is initially independent of DM velocity but hastens once the DM velocity exceeds $v_\x{ion}$.
For a collapsing DM core with a number of particles $N_\x{col}$, the velocity and characteristic collapse time at size $r$ is:
\begin{align}
\label{eq:tcol}
v_\x{col}(r) &\sim \sqrt{\frac{G N_\x{col} m_\chi}{r}}, \nonumber \\
t_\x{col}(r) &\sim \frac{m_\chi v_\x{col}^2}{dE/dt (v_\x{col})} \sim t_2 \min \left [1,~\frac{v_\x{ion}}{v_\x{col}} \right ],
\end{align}
where we have used elastic scatters~\eqref{eq:dEdt} as the dominant dissipation mechanism.
This should be modified once $v_\x{col} \gtrsim 2 \cdot 10^{-2}$ and the momentum transfer becomes $\sim \Lambda_\x{QCD}$.
At this point the interaction is not described by elastic scattering off nuclei, but an inelastic scattering off constituent quarks.
This is a non-perturbative QCD process that will result in the release of pions.
Here the typical energy transfer is $\sim \Lambda_\x{QCD} v_\x{col}$ so the rate of energy loss is instead given by:
\begin{align}
\label{eq:dEdt2}
 \frac{dE}{dt} \sim \Lambda_\x{QCD} n_\x{ion} \sigma_{\chi A} v_\x{col}^2.
\end{align}
We assume that $\sigma_{\chi A}$ is also roughly of order the cross section for this inelastic interaction (with the form factor~\eqref{eq:Helm} set to $A^{-2}$).
At these velocities the collapse time is $t_\x{col} \sim \frac{m_\chi}{n_\x{ion} \sigma_{\chi A} \Lambda_\x{QCD}} \gtrsim 0.5~\x{s}$, with the lower limit assuming a cross section saturating~\eqref{eq:sigma_theta}.
One can also check that $t_\x{col}$ is always greater than the (decreasing) dynamical time $\sim r/v_\x{col}$.

We emphasize that while cooling by nuclear scatters during core collapse is the minimal assumption, other dissipation mechanisms (e.g., radiating as a blackbody) could become efficient due to the increasing DM density, as considered by~\cite{PhysRevD.40.3221}.
However since this is more model-dependent, we do not consider any such additional cooling mechanisms here.

Actually, the initial number of collapsing particles can be parametrically greater than the critical self-gravitation number $N_\x{col} \gg N_\x{sg}$
As discussed in~\cite{Graham:2018efk}, this occurs when the time to capture a self-gravitating number is much less than the time for the DM core to collapse, i.e., when $N_\x{sg} < \Gamma_\x{cap} t_\x{col}$.
We find this is relevant for DM masses $m_\chi \gtrsim 10^{14}~\GeV$.
Here the collapsing core will inevitably ``over-collect" to a much larger number until these two timescales become comparable $N_\x{col} \sim \Gamma_\x{cap} t_\x{col}$, although the density profile of the core at this point is highly non-trivial.
It is worth noting that the collapsing core would likely be non-uniform even in the absence of over-collection, as emphasized in~\cite{PhysRevD.87.123537}---realistically, the core might develop a ``cuspy" profile similar to the formation of galactic DM halos.
In either case, a precise understanding of the DM core density profile is beyond the scope of this work.
For simplicity we will assume a core of \emph{uniform} density with a number of collapsing particles
\begin{equation}
\label{eq:Ncol}
N_\x{col} = \max [ N_\x{sg},~\Gamma_\x{cap} t_\x{col}].
\end{equation}
However, this assumption of a uniform density core is likely a conservative one with regards to our constraints.
For asymmetric DM, a density peak within the collapsing core (e.g. due to over-collection) would collapse to BHs of smaller mass than otherwise assumed and (as we show) would still ignite a SN.
For annihilating DM, a density peak may have a greater rate of annihilations depending on the density profile which would ignite a SN sooner than otherwise assumed.

Though irrelevant prior to self-gravitation, QM effects may become important during the collapse itself.
For a number of collapsing particles $N_\x{col} = N_\x{sg}$, this occurs once the core shrinks within a size:
\begin{equation}
\label{eq:Rqm}
R_\x{QM} \sim \frac{1}{G m_\chi^3 N_\x{sg}^{1/3}} \approx 3 \cdot 10^{-11}~\cm \l\frac{m_\chi}{10^{8} ~\GeV}\r^{-13/6},
\end{equation}
and has a density
\begin{equation}
\label{eq:rhoQM}
\rho_\x{QM} \sim \frac{N_\x{sg} m_\chi}{R_\x{QM}^3} \sim \frac{m_\chi^5 T_\x{WD}^3}{\rho_\x{WD}} \approx 10^{72} \frac{\x{GeV}}{{\cm^3}}\l \frac{m_\chi}{10^{8}~\GeV} \r^5.
\end{equation}
Of course this assumes that the core has not already formed a BH $G N_\x{sg} m_\chi \lesssim R_\x{QM}$.
This means that QM collapse is only relevant for DM masses:
\begin{equation}
\label{eq:qm}
m_\chi \lesssim \frac{\rho_\x{WD}}{T_\x{WD}^3} \approx 10^9~\GeV, ~~~\x{(QM affects collapse),}
\end{equation}
for which it is indeed the case that $N_\x{col} = N_\x{sg}$.
Note that the extreme densities of the DM core~\eqref{eq:rhoQM} are not necessarily problematic as we always assume the DM is point-like with no substructure; however, with an explicit model one should be wary of higher dimension operators modifying the collapse dynamics by potentially triggering new interactions.

\paragraph{Fermionic DM}

If DM is a fermion,~\eqref{eq:Rqm} is precisely the radius of stabilization due to degeneracy pressure.
A degenerate DM core will sit at $R_\x{QM}$ until it collects an additional number of particles $N \gg N_\x{sg}$ and subsequently shrinks as $r \sim \frac{1}{G m_\chi^3 N^{1/3}}$.
Note that additional captured DM particles are still able to dissipate energy and decrease their orbital sizes below the thermal radius under the gravitational influence of the compact core.
For DM masses~\eqref{eq:qm} the collection time $\frac{N}{\Gamma_\x{cap}}$ is always far greater than the cooling time $t_\x{col}$ \eqref{eq:tcol}, and thus the shrinking proceeds adiabatically at a rate $\Gamma_\x{cap}$.

Fermi pressure is capable of supporting a self-gravitating degenerate DM core until it exceeds the Chandrasekhar limit
\begin{equation}
\label{eq:Chafermion}
N^\x{f}_\x{Cha} \sim \frac{\Mpl^3}{m_\chi^3} \approx 2\cdot10^{33} \l \frac{m_\chi}{10^{8} ~\GeV}\r^{-3}.
\end{equation}
%which is of course greater than $N_\x{sg}$ for all DM masses~\eqref{eq:qm}.
Thus the fermi degenerate core will collapse to a BH as long as
\begin{equation}
\label{eq:Chalife}
N^\x{f}_\x{Cha} < N_\x{life}, ~~~~\x{(BH from degenerate core)},
\end{equation}
which is the case for $m_\chi \gtrsim 10^{6}~\GeV$, assuming $\Gamma_\x{cap} = \Gamma_\x{trans}$ and $\rho_\chi = 0.4~\frac{\GeV}{\cm^3}$.
We note that the presence of attractive e.g., Yukawa-type DM self-interactions can drastically reduce the critical number required to overcome Fermi pressure (see~\cite{PhysRevLett.108.191301}), though we do not consider this possibility here.

\paragraph{Bosonic DM}

If DM is a boson, once the DM core collapses to~\eqref{eq:Rqm} it starts populating a BEC.
Further collapse results in increasing the number of particles in the BEC, with the density of the non-condensed particles fixed at $\rho_\x{QM}$, see~\cite{PhysRevD.87.123537} for details.
The size of the BEC is initially set by the gravitational potential of the enveloping self-gravitating sphere, and particles in the BEC have a velocity set by the uncertainty principle:
\begin{align}
\label{eq:becpresg}
r_\x{BEC} &\sim \l \frac{1}{G \rho_\x{QM} m_\chi^2}\r^{1/4} \approx 10^{-16}~\cm \l \frac{m_\chi}{10^{8} ~\GeV} \r^{-7/4}, \nonumber \\
v_\x{BEC} &\sim \frac{1}{m_\chi r_\x{BEC}} \approx 10^{-6} \l \frac{m_\chi}{10^{8} ~\GeV} \r^{3/4}.
\end{align}
The BEC sits at $r_\x{BEC}$ until it becomes self-gravitating at a number:
\begin{equation}
\label{eq:BECsg}
N_\x{BEC, sg} \sim \frac{\rho_\x{QM} r_\x{BEC}^3}{m_\chi} \approx 2\cdot10^{16} \l \frac{m_\chi}{10^{8} ~\GeV} \r^{-5/4}.
\end{equation}
A self-gravitating BEC will continue to add particles, and in the process shrink as $r_\x{BEC} \sim \frac{1}{G m_\chi^3 N}$.
The rate at which DM particles are added to the BEC is set by the rate at which the non-condensed DM core sheds the excess gravitational energy.
The time to condense a number of particles $N \ll N_\x{sg}$ is:
\begin{equation}
\label{eq:tbec}
t_\x{BEC}(N) \sim \frac{N}{N_\x{sg}} t_\x{col}(R_\x{QM}).
\end{equation}
Note that the typical DM velocity in the non-condensed DM sphere at this stage is:
\begin{equation}
\label{eq:vsphere}
v_\x{col}(R_\x{QM}) \sim \sqrt{\frac{G N_\x{sg} m_\chi}{R_\x{QM}}} \approx 0.3~\l\frac{m_\chi}{10^{8}~\GeV}\r^{1/3}.
\end{equation}

The pressure induced by the uncertainty principle is capable of supporting the self-gravitating sphere of DM particles until it exceeds the so-called bosonic Chandrasekhar limit:
\begin{equation}
\label{eq:boscha}
N^\x{b}_\x{Cha} \sim \frac{\Mpl^2}{m_\chi^2} \approx 10^{22} \l \frac{m_\chi}{10^{8} ~\GeV}\r^{-2},
\end{equation}
which is far less than $N_\x{sg}$ for all DM masses~\eqref{eq:qm}.
Interestingly, this limit is dramatically affected by even the presence of miniscule DM self-interactions~\cite{PhysRevLett.57.2485}.
These may be a generic expectation given the already assumed scattering cross section off nucleon, as emphasized in~\cite{PhysRevD.87.123507}.
In the case of a repulsive $\lambda |\chi|^4$ interaction potential where $\lambda>0$, no stable configuration exists beyond a critical number
\begin{equation}
N^\x{b}_\x{Cha, self} \sim \frac{\Mpl^2}{m_\chi^2}  \l 1+ \frac{\lambda}{32 \pi} \frac{\Mpl^2}{m_\chi^2}\r^{1/2}.
\end{equation}
We find that $N^\x{b}_\x{Cha, self}$ is still less than $N_\x{sg}$ as long as $\lambda \lesssim 10^{-2}$.
An attractive self-interaction could reduce the necessary critical limit, although this is highly model-dependent.
From here on, we will use~\eqref{eq:boscha} as the relevant critical limit.

\subsection{Annihilating DM Collapse}
\label{sec:AnnihilatingDM}

Now consider the case of DM with an annihilation cross section $\sigma_{\chi \chi}$ into SM products, e.g., quarks.
We will restrict our attention here to DM masses $m_\chi \ll \Eboom$ such that multiple annihilations are necessary to ignite a SN.
As in the asymmetric case, for simplicity we focus on DM which scatters infrequently, $\sigma_{\chi A} < \sigma_\theta$.

As described above, the thermalizing DM constitutes a number density of DM throughout the WD volume.
Depletion of this in-falling DM is dominated by the total rate of annihilations near the thermal radius:
\begin{equation}
\Gamma_\x{infall} \sim \frac{(\Gamma_\x{cap} t_2)^2}{R_\x{th}^3} \sigma_{\chi \chi} v_\x{th}.
\end{equation}
Therefore a DM core at $R_\x{th}$ will steadily collect at a rate roughly $\Gamma_\x{cap}$ as long as
\begin{equation}
\label{eq:steadycollect}
\Gamma_\x{infall} < \Gamma_\x{cap}, ~~~\x{(steady DM collection)}.
\end{equation}
Of course this collecting DM core is also depleting via annihilations, and will at most reach an equilibrium number
\begin{align}
\label{eq:Neq}
N_\x{eq} \sim \l \frac{\Gamma_\x{cap} R_\x{th}^3}{\sigma_{\chi \chi} v_\x{th}} \r^{1/2}.
\end{align}
This results in a more stringent condition for self-gravitation:
\begin{equation}
\label{eq:sgcondann}
N_\x{sg} < \x{min} [N_\x{life}, N_\x{eq} ], ~~~\x{(core self-gravitates)}.
\end{equation}
If $N_\x{sg} > N_\x{life}$ or $N_\x{sg} > N_\x{eq}$, the DM core has either saturated at a number $N_\x{eq}$ or is still continuing to collect at a number $N_\x{life}$, whichever comes first.
In either case if the core does not reach self-gravitation (i.e.~\eqref{eq:sgcondann} is not satisfied), we found that the total rate of annihilations within a core subregion of volume $\lambda_T^3 \ll R_\x{th}^3$ is much too small to ignite a SN.

\begin{figure}
\centering
\includegraphics[width=8cm]{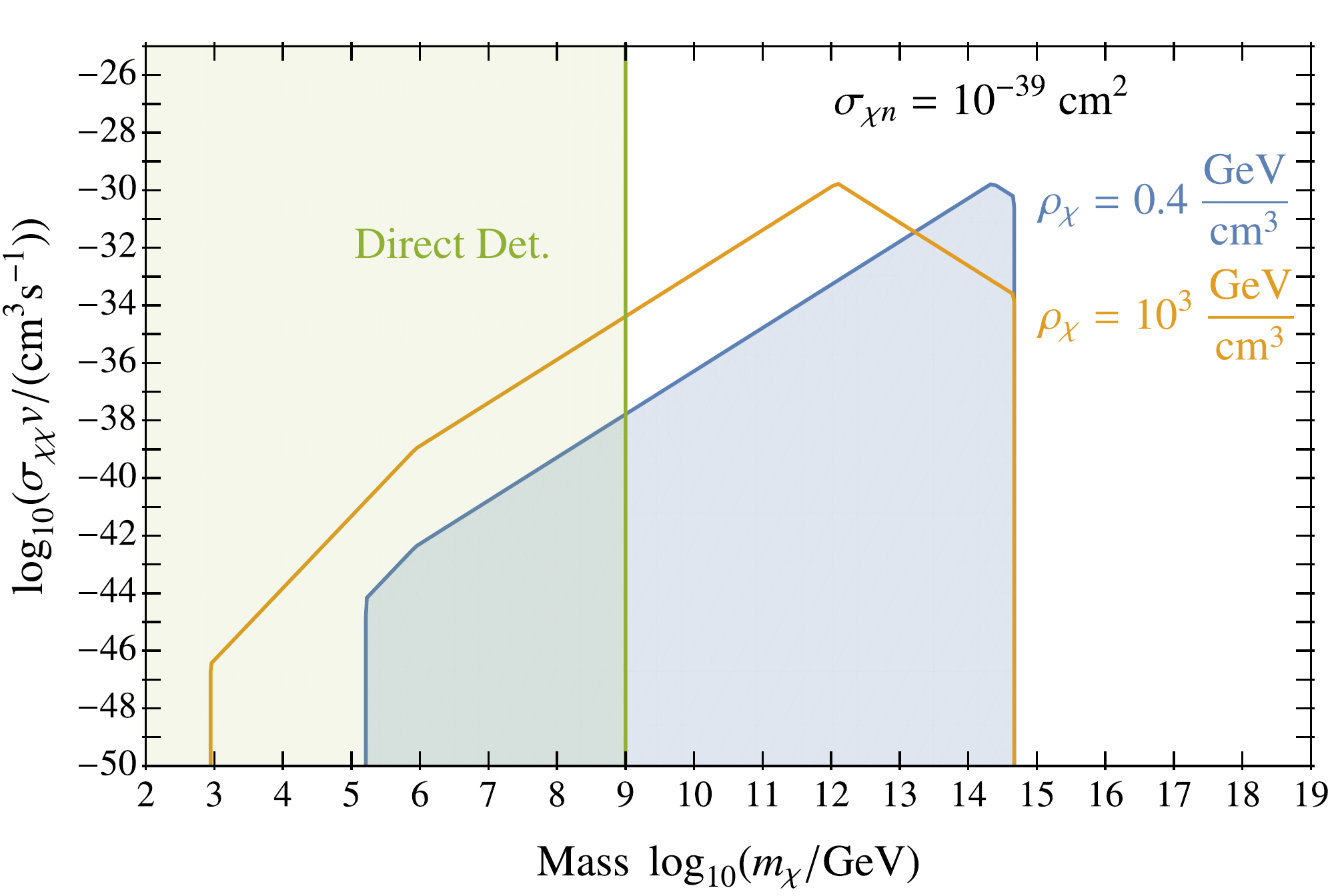}
\caption{Parameter space $\{ m_\chi, \sigma_{\chi \chi} v\}$ of annihilating DM in which a DM core forms and collapses within $\tau_\x{WD} \approx 5~\x{Gyr}$ in a WD of local DM density $\rho_\chi$.
We take a fixed value of the DM-nuclei scattering cross section $\sigma_{\chi n} = 10^{-39}~\cm^2$.
See text for details.
}
\label{fig:collapsecondann}
\end{figure}

We thus turn to core collapse, during which annihilations become more rapid as the core shrinks.
The conditions~\eqref{eq:core},~\eqref{eq:steadycollect} and~\eqref{eq:sgcondann} on the $\{m_\chi ,\sigma_{\chi \chi} v \}$ parameter space for which a collapse takes place are depicted in Fig.~\ref{fig:collapsecondann}.
Here we have taken a fixed fiducial value of the scattering cross section $\sigma_{\chi n} = 10^{-39}~\cm^2$, though the allowed parameter space of collapse in the case of annihilating DM exists for any $\sigma_{\chi n}$ within the region shown in Fig.~\ref{fig:collapsecond}.
We have checked that there are no existing constraints at these low DM annihilation cross sections, for instance from DM annihilations in the galactic halo contributing to the observed cosmic ray flux.

As before, a self-gravitating DM core shrinks at a rate set by cooling~\eqref{eq:tcol}.
However the core is also annihilating so that $N(r)$ is decreasing from its initial value $N_\x{col}$~\eqref{eq:Ncol}.
When the DM core is at a radius $r$, the total rate of annihilations is:
\begin{equation}
\label{eq:Gammacol}
\Gamma_{\chi \chi} \sim \frac{N^2}{r^3}\, \sigma_{\chi \chi} v_\x{col},
\end{equation}
The collapse will initially proceed unscathed, with the number of collapsing particles roughly constant $N(r) \approx N_\x{col}$, until the characteristic annihilation time $\frac{N}{\Gamma_{\chi \chi}}$ is of order the collapse time $t_\x{col}$.
The size of the core at this stage is an important scale, which we denote as $R_{\chi \chi}$.
Note that $R_{\chi \chi}$ as defined is trivially smaller than $R_\x{th}$ if conditions~\eqref{eq:steadycollect} and~\eqref{eq:sgcondann} are satisfied.
The expression for $R_{\chi \chi}$ depends on whether this takes place during the viscous or inertial drag regimes, or in the inelastic scattering regime~\eqref{eq:dEdt2}.
Written in terms of the annihilation cross-section $\sigma_{\chi \chi} v_\x{col}$, this scales as:
\begin{align}
\label{eq:Rchichi}
R_{\chi \chi} \propto
\begin{cases}
 &(\sigma_{\chi \chi} v_\x{col})^{1/3} ~~~~ v_\x{col} < v_\x{ion} ~ \x{or} ~2\cdot 10^{-2} < v_\x{col}\\
 &(\sigma_{\chi \chi} v_\x{col})^{2/5} ~~~~  v_\x{ion} < v_\x{col} < 2\cdot 10^{-2} \\
 \end{cases}.
\end{align}
Note that $v_\text{col}$ is to be evaluated at $R_{\chi\chi}$ in these expressions.

Once the DM core collapses to within $R_{\chi \chi}$, it begins depleting appreciably.
We call this an annihilation burst.
Once $r \lesssim R_{\chi \chi}$, the continued evolution of the DM core is driven by two competing effects:
scatters with the stellar matter drive the core to collapse to smaller radii, as before, but at the same time annihilations drive the core to expand by weakening the gravitational potential.
We do not work out this detailed evolution, but rather conservatively consider the constraints only for $r \gtrsim R_{\chi \chi}$.

For DM masses~\eqref{eq:qm}, if $R_{\chi \chi} > R_\x{QM}$ then the core effectively annihilates before any quantum statistics become significant.
On the other hand, if $R_{\chi \chi} < R_\x{QM}$ then the core remains roughly intact and can form a fermi degenerate core or BEC, as in the asymmetric DM case.
We examine the subsequent evolution of the core in the case $R_{\chi \chi} < R_\x{QM}$, but with the added presence of annihilations.

\paragraph{Fermionic DM}

If DM is a fermion, a fermi degenerate core will continue to collect DM particles and shrink (and thus the rate of annihilations increases).
During this stage, the degenerate DM core can saturate at an equilibrium $N^\x{f}_{\chi \chi}$ when the annihilation rate $\Gamma_{\chi \chi}$ is of order the shrinking rate set by DM capture $\Gamma_\x{cap}$.
If $N^\x{f}_{\chi \chi} \lesssim N_\x{sg}$, the fermi degenerate core saturates while still roughly at $R_\x{QM}$~\eqref{eq:Rqm}.
If $N^\x{f}_{\chi \chi} \gtrsim N_\x{sg}$, the core substantially shrinks before saturating at a number:
\begin{equation}
\label{eq:Nchichif}
N^\x{f}_{\chi \chi} \sim \frac{\Gamma_\x{cap}^{1/3}}{G m_\chi^3 (\sigma_{\chi \chi} v_\x{col})^{1/3}}, ~~~ N^\x{f}_{\chi \chi} > N_\x{sg}.
\end{equation}
Of course, for sufficiently low annihilation cross section a saturated core may never form in the WD lifetime $N_\x{life} < N^\x{f}_{\chi \chi}$ or before forming a BH $N^\x{f}_\x{Cha} < N^\x{f}_{\chi \chi}$.

\paragraph{Bosonic DM}

If DM is a boson the core will condense particles into a BEC.
As the non-condensed core collapse proceeds at constant density, it will never burst as the rate of annihilations in the enveloping sphere only decreases.
However the BEC can saturate at an equilibrium number when the annihilation rate in the compact region becomes of order the condensation rate given by~\eqref{eq:tbec}.
We have checked that this saturation is never reached before the BEC self-gravitates at a number~\eqref{eq:BECsg}.
Subsequently the BEC adds particles from the core and shrinks (and the rate of annihilations in the BEC increases).
The self-gravitating BEC then either saturates at a number
\begin{align}
\label{eq:Nchichib}
N^\x{b}_{\chi \chi} \sim \l \frac{N_\x{sg}}{t_\x{col}(R_\x{QM}) G^3 m_\chi^{9} \sigma_{\chi \chi} v_\x{BEC}} \r^{1/5}, ~~ N^\x{b}_{\chi \chi} > N_\x{BEC, sg}.
\end{align}
or first reaches $N^\x{b}_\x{Cha}$ when annihilations are negligible and forms a BH.

\subsection{Endgame}
\label{sec:endgame}

There are many possible outcomes of the DM core collapse in a WD.%
\footnote{The number of possible outcomes may be $\geq 14,000,605$ \cite{strange}.}
For asymmetric DM the core can collapse to a mini BH, either directly or by first forming a fermi degenerate core or populating a BEC.%
\footnote{This can only take place after entering the Quantum Realm.}
As detailed in Sec.~\ref{sec:BH}, such a BH can ignite a SN by emission of Hawking radiation or, as we motivate, possibly even during its accretion.
For annihilating DM the core annihilates at an increasing rate until collapsing to $R_{\chi \chi}$, at which point it is effectively annihilating an $\OO(1)$ fraction.
As detailed in Sec.~\ref{sec:annburst}, this large number of rapid annihilations can even ignite a SN before the core reaches $R_{\chi \chi}$.

It is also the case that the DM core is directly heating the WD via nuclear scatters.
This may be sufficient to ignite a SN, as first calculated by~\cite{Bramante:2015cua}.
We estimate the total energy deposited by a collapsing core of size $r$ inside a trigger region $\lambda_T^3$ during a time $\tau_\x{diff}$ as:
\begin{align}
\label{eq:Escat}
\mathcal{E}_{\chi A} (r) \sim N_\x{col} m_\chi v_\x{col}^2 \l \frac{\tau_\x{diff}}{t_\x{col}} \r \cdot \x{min} \left[1, \left(\frac{\lambda_T}{r} \right)^3\right].
\end{align}

In considering this process,~\cite{Bramante:2015cua} additionally required that (1) the DM core be self-thermalized (e.g., due to DM-DM self interactions)
%with a cross section of order the DM-nucleon scattering cross section
and (2) the core must uniformly heat a trigger region $\lambda_T^3$, thus restricting the analysis to core sizes $r \gtrsim \lambda_T$.
Neither of these requirements are necessary, however.
While a deposited energy well inside the trigger region may not immediately ignite a conductive flame as per~\cite{woosley}, it will eventually if the energy is sufficiently large~\eqref{eq:boom2} once the heat has diffused out to a size $\sim \lambda_T$ (see~\cite{Graham:2018efk} for a more detailed discussion of this evolution).
This observation allows the derived constraints of~\cite{Bramante:2015cua} to be extended to larger DM masses: we simply require $\mathcal{E}_{\chi A} \gtrsim \Eboom$ satisfies the condition~\eqref{eq:boom2} in order for scattering to ignite a SN.

We emphasize that the heat deposited in the stellar matter during a DM collapse would be drastically affected by the presence of an additional cooling mechanism which drives the collapse, e.g., emitting dark radiation.
In particular, if such a cooling mechanism is present and efficient in a collapsing core, ignition due to heating by nuclear scatters as in~\cite{Bramante:2015cua} might not occur.
As we show in Sec.~\ref{sec:BH} and Sec.~\ref{sec:annburst}, however, most collapsing DM cores would still ignite a SN from BH formation or annihilations.
For this reason, while we show the extended constraints on DM-nuclear scatters from~\eqref{eq:Escat}, we will also consider and show the consequences of core collapse to smaller radii, below the size at which nuclear scatters (as the sole cooling mechanism) would deposit sufficient energy to be constrained.

\section{Black hole-induced SN}
\label{sec:BH}

\begin{figure}
\label{fig:BHmass}
\centering
\includegraphics[width=8cm]{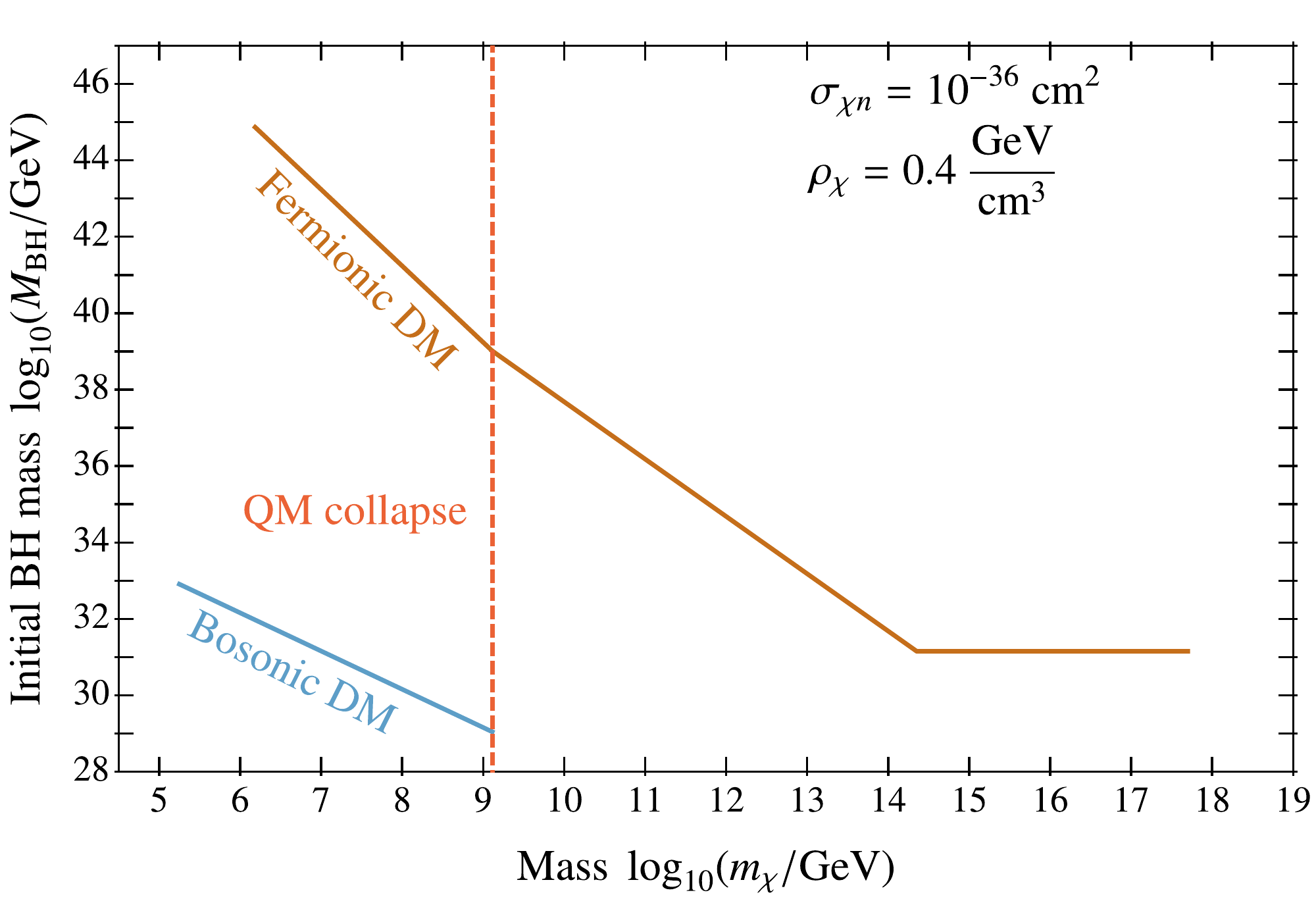}
\caption{Initial black hole mass formed by DM core collapse in a WD.
We take a representative value of the scattering cross section, though $M_\x{BH}$ is independent of $\sigma_{\chi n}$ except for large DM masses where $N_\x{sg} < \Gamma_\x{cap} t_\x{col}$.
As plotted $M_\x{BH}$ cuts-off at points where a DM core does not even form or collapse, or where a fermi degenerate core does not have time to collect a Chandrasekhar number $N_\x{Cha}^\x{f}$.
}
\end{figure}

As described in Sec.~\ref{sec:AsymmetricDM}, a BH formed by DM collapse will have an initial mass (shown in Fig.~\ref{fig:BHmass}):
\begin{equation}
M_\x{BH} \sim
\begin{cases}
N^\x{f}_\x{Cha} m_\chi ~~~~~ m_\chi \lesssim 10^9~\GeV ~~~ \x{fermionic DM} \\
N^\x{b}_\x{Cha} m_\chi ~~~~~ m_\chi \lesssim 10^9~\GeV ~~~ \x{bosonic DM} \\
G N_\x{col} m_\chi ~~~~ m_\chi \gtrsim 10^{9}~\GeV \\
\end{cases}.
\end{equation}
%Here $N_\x{col}$ is given by~\eqref{eq:Ncol}.
Note that any such BH will necessarily have some angular momentum.
The DM core initially inherits its angular velocity from the rotating WD, though loses angular momentum to the stellar medium as it cools and collapses.
We find the dimensionless spin parameter of the initial BH is always small $\frac{J_\x{BH}}{G M_\x{BH}^2} \lesssim 10^{-2}$, assuming a WD angular velocity of $\Omega_\x{WD} \approx 0.01~\text{Hz}$.
Thus the newly formed BH is approximately Schwarzschild, and has a radius:
\begin{equation}
R_\x{BH} = 2 G M_\x{BH} \approx 3 \cdot 10^{-5}~\cm \l \frac{M_\x{BH}}{10^{47}~\GeV} \r.
\end{equation}

\subsection{Fate of a BH}

It is generally believed~\cite{Hawking:1974sw} that BHs have a temperature
\begin{equation}
T_\x{BH} = \frac{1}{4 \pi R_\x{BH}} \approx 6~\MeV \l \frac{M_\x{BH}}{10^{39}~\GeV} \r^{-1},
\end{equation}
and lose mass by emitting particles at a rate
\begin{equation}
\label{eq:hawking}
\l \frac{d M_\x{BH}}{dt} \r_\x{HR}= \frac{\alpha}{G^2 M_\x{BH}^2},
\end{equation}
where $\alpha(M_\x{BH})$ encodes the different particle emission rates, roughly increasing as the BH temperature exceeds the mass threshold of a new species.
Detailed calculation~\cite{Page:1976df} finds $\alpha \approx 2.8 \cdot 10^{-4}$ for $T_\x{BH} \lesssim \MeV$, accounting for emission of photons, gravitons, and three neutrino species.
Counting only experimentally verified SM degrees of freedom, the emission rate effectively asymptotes to $\alpha \approx 4.1 \cdot 10^{-3}$ for $T_\x{BH} \gtrsim 100~\text{GeV}$~\cite{Ukwatta:2015iba}.
%For simplicity we approximate $\alpha(M)$ within this mass range as a simple power law.
Thus an evaporating BH (by this we mean a BH which only Hawking radiates without any accretion)%
\footnote{An evaporating BH loses angular momentum rapidly and has a decreasing spin parameter---thus rotation is negligible throughout the evaporation~\cite{PhysRevD.14.3260}.
}
has a lifetime less than $\tau_\x{WD} \approx 5~\x{Gyr}$ if:
\begin{equation}
\label{eq:HRlife}
M_\x{BH} \lesssim 2 \cdot 10^{38} ~\GeV~~~ (\x{evaporate in $\tau_\x{WD}$}).
\end{equation}

The BH primarily accretes nuclear matter and additional DM particles: which dominates depends on the BH mass, or more precisely the DM parameters.
In the hydrodynamic spherical so-called Bondi approximation, the former is given by
\begin{equation}
\label{eq:bondi}
\l \frac{d M_\x{BH}}{dt} \r_\x{WD} = 4 \pi \lambda \l  \frac{G M_\x{BH}}{c_s^2} \r^2 \rho_\x{WD} c_s,
\end{equation}
where $c_s \approx 2 \cdot 10^{-2}$ is the sound speed (approximated from numerical calculations in~\cite{Balberg:2000xu}), and $\lambda \sim \OO(1)$~\cite{shapiro}.

The accretion of DM potentially has two contributions.
Under the influence of the BH gravitational potential, individual DM particles will continue reducing their orbit size below the thermal radius by scattering with the stellar medium.
Once it crosses the angular momentum barrier $4 G M_\x{BH}$, the DM will rapidly fall into the BH~\cite{shapiro}.
A steady state is soon achieved after the BH is formed where DM feeds the BH at a rate set by the capture rate:
\begin{equation}
\label{eq:steadyDM}
\l \frac{d M_\x{BH}}{dt} \r_\chi= \Gamma_\x{cap} m_\chi
\end{equation}
There may also be large overdensity of DM particles in the vicinity of the newly formed BH, which is likely if the DM core collapses with non-uniform density.
In the collisionless spherical approximation~\cite{shapiro}, a DM population with density $\rho_\infty$ and velocity $v_\infty$ far from the BH accretes at a rate:
\begin{equation}
\label{eq:overdenseDM}
\l \frac{d M_\x{BH}}{dt} \r_\chi = \frac{16 \pi \rho_\infty G^2 M_\x{BH}^2}{v_\infty}.
\end{equation}
Such accretion is especially relevant for bosonic DM if the BH is formed from a compact BEC within an enveloping non-condensed DM core~\cite{PhysRevD.87.123537}.
For our purposes we will only consider~\eqref{eq:overdenseDM} in this scenario, where $\rho_\infty$ is given by the very large density~\eqref{eq:rhoQM} and $v_\infty$ is given by~\eqref{eq:vsphere}.

The fate of a BH is determined by:
\begin{align}
\label{eq:BHevol}
\frac{d M_\x{BH}}{dt} = &-\l \frac{d M_\x{BH}}{dt} \r_\x{HR}\nonumber \\
&+ \l \frac{d M_\x{BH}}{dt} \r_\x{WD}+ \l \frac{d M_\x{BH}}{dt} \r_\chi.
\end{align}
We first consider BHs that are not formed from a BEC.
Without DM accretion, we find Hawking evaporation beats Bondi accretion, i.e., $\l \frac{d M_\x{BH}}{dt} \r_\x{HR} > \l \frac{d M_\x{BH}}{dt} \r_\x{WD}$  at masses:
\begin{equation}
\label{eq:Hawkingbeatsbondi}
M_\x{BH} \lesssim 10^{38}~\GeV,~~~~\x{(Hawking beats Bondi)}.
\end{equation}
Including the steady accretion of DM~\eqref{eq:steadyDM}, we find Hawking evaporation beats the largest possible DM accretion, i.e., $\l \frac{d M_\x{BH}}{dt} \r_\x{HR} > \l \frac{d M_\x{BH}}{dt} \r_\chi$ when $\Gamma_\x{cap} = \Gamma_\x{trans}$ at masses
\begin{equation}
\label{eq:HawkingbeatsDM}
M_\x{BH} \lesssim 2\cdot10^{35}~\GeV,~~~~\x{(Hawking beats DM)},
\end{equation}
where Hawking also clearly beats Bondi.
$M_\x{crit}$ depends on the strength of the steady DM accretion~\eqref{eq:steadyDM}, and for the relevant DM parameter space lies in the range:
\begin{equation}
\label{eq:Mcrit}
M_\x{crit} \approx 2\cdot 10^{35} - 10^{38}~\GeV.
\end{equation}
where the upper end of this range holds when Bondi dominates the accretion, and all lower values apply when steady DM accretion~\eqref{eq:steadyDM} dominates.

We now consider the timescales involved in accreting or evaporating, which can estimated by the characteristic time:
\begin{equation}
\tau_\x{BH} \sim \frac{M_\x{BH}}{d M_\x{BH}/dt}.
\end{equation}
If the BH is evaporating, $\tau_\x{BH} \propto M_\x{BH}^3$ and is set by the time spent at the largest BH mass, i.e. the initial BH mass.
If the BH is dominantly accreting by Bondi then $\tau_\x{BH} \propto M_\x{BH}^{-1}$ is set by the time spent at the smallest BH mass,
If, however, the BH is dominantly accreting by DM~\eqref{eq:steadyDM} then $\tau_\x{BH} \propto M_\x{BH}$ is instead set by the time spent at the largest BH mass---this is the BH mass at which Bondi accretion takes over $10^{38} \lesssim M_\x{BH} \lesssim 10^{41}~\GeV$ (depending on the capture rate $\Gamma_\x{cap}$).
Miraculously, we find $\tau_\x{BH} \approx \x{Gyr}$ for BH masses $M_\x{BH} \approx 10^{38}~\GeV$, coinciding with the upper end of~\eqref{eq:Mcrit} where Bondi accretion becomes of order the Hawking evaporation.
This can also be seen from the fact that $M_\x{crit}$~\eqref{eq:Mcrit} lies \emph{just} below the BH mass necessary to evaporate within $\tau_\x{WD} \approx 5~\x{Gyr}$ in the absence of any accretion~\eqref{eq:HRlife}.
Thus it is clear that whether the BH is evaporating or accreting, it will necessarily do so in a characteristic time less than a Gyr.

Returning to the case of BHs formed from a BEC, we find that the DM accretion of the non-condensed enveloping DM core~\eqref{eq:overdenseDM} in fact beats Hawking evaporation over the entire DM mass range of interest.
Note that this outcome is strikingly different from the analogous process in a NS, where it has been found that such BHs always dominantly evaporate \cite{PhysRevD.87.123537}.
The difference arises from the fact that the density of the DM core~\eqref{eq:rhoQM} is significantly smaller at NS densities/temperatures and at the lower DM masses considered by~\cite{PhysRevD.87.123537}.

We now briefly address the question: is Bondi always a valid estimate for the accretion of nuclear matter onto the BH?
As is well-known, accretion could be in the Eddington-limited regime: this occurs when the radiation produced by in-falling matter exerts a significant pressure so as to back-react on the accretion.
In the spherical approximation, this yields a maximum luminosity:
\begin{equation}
\label{eq:eddington}
L_\x{edd} = \frac{4 \pi G M_\x{BH} m_\x{ion}}{\sigma},
\end{equation}
where $\sigma$ is the dominant interaction by which outgoing radiation transfers momentum to the in-falling matter.
Assuming photon energies near the horizon $\omega \gtrsim~\MeV$, this is either set by hard Compton scattering off electrons $\sigma~\sim~\frac{\alpha^2}{m_e \omega} \sim 100~\x{mb} \l \frac{\omega}{\MeV} \r^{-1}$ or inelastic photo-nuclear interactions off ions $\sigma \sim \x{mb}$ (see~\cite{Graham:2018efk} for details).
Accretion is Eddington-limited if $\epsilon \cdot (d M_\x{BH}/dt)_\x{WD}$ exceeds $L_\x{edd}$, where $\epsilon$ is the radiation efficiency.
If we conservatively take $\epsilon \sim 0.1$, we find Bondi accretion is not Eddington-limited for BH masses less than $M_\x{BH} \lesssim 10^{40}~\GeV$.
Note that even if the accretion is Eddington-limited at larger BH masses, the timescale $\tau_\x{BH}$ then becomes independent of $M_\x{BH}$ and is still much less than a Gyr.

The accretion could also be stalled by the stellar rotation: this occurs when the in-falling matter possesses excess angular momentum that must be dissipated to accrete, e.g., by viscous stresses during a slow phase of disk accretion~\cite{shapiro}.
\cite{Kouvaris:2013kra} examines the effect of rotations for mini BHs in NSs, concluding that kinematic viscosity can maintain Bondi spherical accretion as long as the BH mass is sufficiently small.
Based on the analysis of~\cite{Kouvaris:2013kra}, we crudely estimate that Bondi accretion would hold for $M_\x{BH} \lesssim 10^{46}~\GeV$, assuming a (conservative choice of) WD viscosity~\cite{DallOsso:2013uac}.
Even if the BH accretion is stalled beyond this point we suspect the accretion timescale is still much smaller than a Gyr, though a detailed understanding is beyond the scope of this work.

\subsection{Constraints}

\paragraph{Hawking.}
The Hawking radiation emitted by a BH will ignite a SN if
\begin{equation}
\label{eq:Hawkingboom}
\mathcal{E}_\x{BH} \sim \frac{\alpha}{G^2 M_\x{BH}^2} \cdot \min [\tau_\x{diff}, \tau_\x{BH}]
\end{equation}
satisfies the condition~\eqref{eq:boom2} $\mathcal{E}_\x{BH} \gtrsim \Eboom$.
If the BH is evaporating, then $\tau_\x{BH}$ is just its remaining lifetime (which is greater than $\tau_\x{diff}$ for BH masses $M_\x{BH} \gtrsim 10^{29}~\GeV$).
Even if a BH is technically accreting, it is possible to ignite a SN by the large amount of Hawking radiation emitted during its infancy.
In this case, one can check that~\eqref{eq:Hawkingboom} still approximates the dominant contribution to the total energy emitted during a time $\tau_\x{diff}$.
%This is because we only consider accretion processes scaling as a non-negative power of the BH mass $\dot{M} \propto M^n$, where $n \geq 0$, and thus Hawking radiation is dominantly emitted at the lowest masses.

Assuming $\tau_\x{diff} \ll \tau_\x{BH}$, applicable for all starting BH masses we consider, Hawking is explosive at BH masses:
\begin{equation}
\label{eq:HawkingM}
M_\x{BH, boom} \approx 2\cdot10^{35}~\GeV.
\end{equation}
Of course, any DM core that results in a BH initially less than $M_\x{BH, boom}$ ignites a SN upon formation.
In addition, DM cores that result in a BH initially greater than $M_\x{BH, boom}$ but less than the critical threshold $M_\x{crit}$ evaporate and eventually ignite a SN within a Gyr.
Coincidentally, any BH initially greater than $M_\x{crit}$ will not ignite a SN via Hawking but will instead accrete---this is evident from the fact that~\eqref{eq:HawkingM} lies just below the lower end of the critical threshold~\eqref{eq:Mcrit}.
However this is notably not the case for accreting BHs formed from a BEC: we have checked that all BHs formed from a BEC \emph{immediately} ignite a SN by Hawking despite the large accretion rate from the large enveloping DM density.

\paragraph{Accretion.}
Finally, we comment on the final outcome of an accreting BH.
It is conservative to suppose that such a BH simply eats the star.
However, it is plausible that accreting BHs in WDs ignite SN once they grow sufficiently large.
We can think of at least two potential mechanisms for this:

(1)
The flow of stellar matter into the BH leads to the formation of a sonic horizon $R_s \sim G M_\x{BH}/c_s^2 \sim 10^4~R_\x{BH}$, with supersonic flow as the matter enters free-fall near the BH.
The kinetic energy of a carbon ion at the sonic horizon is $m_\text{ion}c_s^2 \sim \text{MeV}$, increasing as it falls inward.
It is reasonable to suppose that the flow inside the sonic horizon is not perfectly radial, in which case this violent swarm of carbon ions may ignite thermonuclear fusion.
BH masses $M_\x{BH} \gtrsim 10^{43}~\GeV$ have sonic horizons $R_s \gtrsim \lambda_T$.
Assuming substantial non-radial flow, such BHs may then have carbon ions colliding at large enough energies to overcome the coulomb barrier and initiate fusion over a large region.
As this fusion is happening within the sonic horizon, a resulting fusion front would need to propagate out as a supersonic shockwave (e.g., a so-called detonation front~\cite{kippenhan}) in order to ignite the rest of the star.

(2)
Inflow onto the BH also increases the density of stellar matter near the BH, for instance by roughly a factor $ \sim 10-100$ at the sonic horizon~\cite{shapiro}.
This increased density may be sufficient, even at low temperatures, to ignite the star outside the sonic horizon through pycnonuclear fusion without the need for a supersonic shockwave
(or inside the sonic horizon, with an accompanying supersonic fusion front.)
Runaway pycnonuclear fusion begins when a sufficiently large region of carbon achieves a critical density $\sim 10^{10}~\text{g/cm}^3$~\cite{kippenhan}, which is a factor $ \sim 30$ greater than our chosen central density.
Note that the corresponding pycnonuclear trigger size $\lambda_P$ may be different from the thermonuclear trigger size $\lambda_T$ as the rates of fusion and diffusion depend on density and temperature, and both may be modified by dynamics near the BH.
However, if we simply assume $\lambda_P \sim \lambda_T \sim 10^{-5}~\cm$, then large BH masses $M_\x{BH} \gtrsim 10^{44}~\text{GeV}$ would have a sonic horizon $R_s \gg \lambda_P$, and could thus potentially ignite a SN via subsonic fusion front.

To confirm either of these mechanisms leads to ignition would require more detailed numerical calculations, which we do not attempt here.
In any case, whether an accreting BH eats the star or ignites a SN, we are able to constrain any such BHs by the existence of observed WDs given that the accretion timescale is less than a Gyr.

\begin{figure}
\centering
\includegraphics[width=8cm]{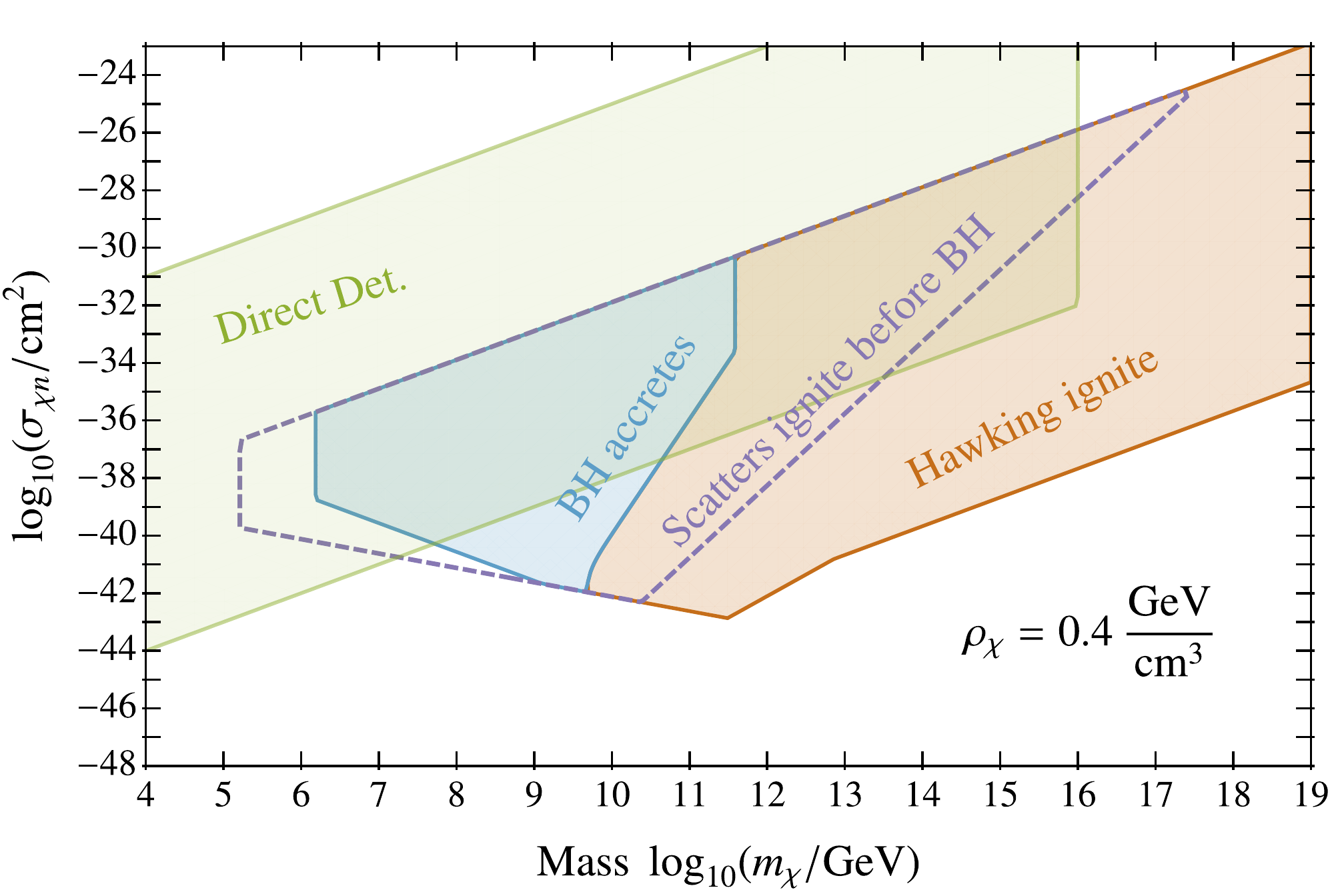}
\caption{Constraints on fermionic asymmetric DM which forms a DM core and collapses to a mini black hole in a WD.
The black hole either ignites a supernova via Hawking emission ({\color{c6} red}) or accretes and eats the star (or possibly ignites a supernova) ({\color{c7} blue}).
Also shown ({\color{c5} purple}) are the constraints on DM-nuclei scatters igniting a supernova during core collapse before formation of a black hole.
}
\label{fig:BHfermion}
\end{figure}

\begin{figure}
\centering
\includegraphics[width=8cm]{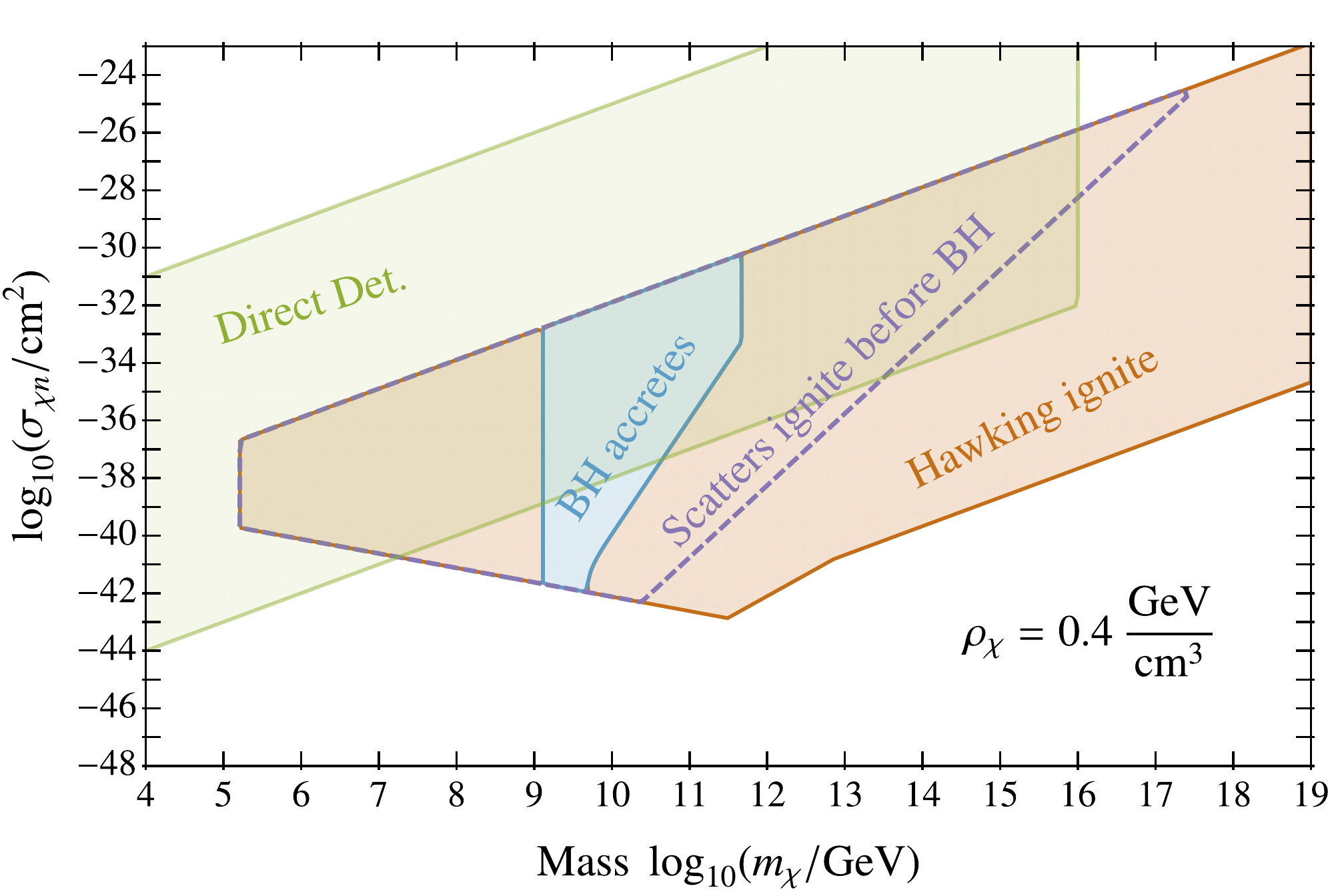}
\caption{Constraints on bosonic asymmetric DM which forms a DM core and collapses to a mini black hole in a WD.
The black hole either ignites a supernova via Hawking emission ({\color{c6} red}) or accretes and eats the star (or possibly ignites a supernova) ({\color{c7} blue}).
Also shown ({\color{c5} purple}) are the constraints on DM-nuclei scatters igniting a supernova during core collapse before formation of a black hole.
}
\label{fig:BHboson}
\end{figure}

To summarize, BHs formed by DM core collapse will either ignite a SN by Hawking radiation, or accrete and subsequently eat the star or ignite a SN.
The resulting constraints on DM parameters are shown in Fig.~\ref{fig:BHfermion} (fermionic DM) and Fig.~\ref{fig:BHboson} (bosonic DM).
For fermionic DM these constraints extend well beyond those previously derived which consider BH formation/accretion in NSs, and are thus complementary.
For bosonic DM these constraints are entirely new---in the DM mass range of interest, there are in fact no bounds due to BH formation in NSs (see~\cite{PhysRevD.87.123537} for details).
We also show the constraints from DM-nuclei scatters igniting a SN during core collapse at any point before formation of a BH (or a fermi degenerate core or BEC).

\section{Annihilation-induced SN}
\label{sec:annburst}

A collapsing core of annihilating DM has an increasing annihilation rate, and effectively depletes $\OO(1)$ (``bursts") upon shrinking to a size $r \sim R_{\chi \chi}$.
However, even while $r \gtrsim R_{\chi \chi}$ and the DM core roughly retains its initial number $N(r) \approx N_\x{col}$, the energy deposited by a small fraction of the core may be significant.
We estimate the energy deposited in the large number of annihilations within a trigger region $\lambda_T^3$ and diffusion time $\tau_\x{diff}$ for $r \gtrsim R_{\chi \chi}$:
\begin{align}
\label{eq:Echichi}
\mathcal{E}_{\chi \chi}(r) \sim m_\chi \frac{N_\x{col}^2}{r^3} \sigma_{\chi \chi} v_\x{col} \tau_\x{diff} \cdot \x{min} \left[1, \left(\frac{\lambda_T}{r} \right)^3\right].
\end{align}
This is sufficient to ignite a SN if it satisfies $\mathcal{E}_{\chi \chi} \gtrsim \Eboom$~\eqref{eq:boom2}.

As expected, the annihilating core deposits energy more and more rapidly as it shrinks to smaller radii.
We can also evaluate the deposited energy~\eqref{eq:Echichi} at the bursting point $r \sim R_{\chi \chi}$.
Interestingly, $\mathcal{E}_{\chi \chi}(R_{\chi \chi})$ scales inversely with annihilation cross section $\mathcal{E}_{\chi \chi}(R_{\chi \chi}) \propto (\sigma_{\chi \chi} v_\x{col})^{-1/5}$ in the regime $v_\x{ion} < v_\x{col}(R_{\chi \chi}) < 2\cdot 10^{-2}$, i.e. the DM core is more explosive for \emph{lower} annihilation cross section.
This is basically a result of the collapsing core focusing and becoming more dense before annihilating $\OO(1)$, thus making this energy deposition at $r \sim R_{\chi \chi}$ more violent.
It is also interesting that $\mathcal{E}_{\chi \chi}(R_{\chi \chi})$ scales inversely with DM mass---this is just a result of the greater number of collapsing particles at lower DM masses.

If the core has not yet ignited a SN by the time it collapses to $R_{\chi \chi}$, could it do so afterwards?
Although the number of collapsing particles at this point is depleting appreciably, the shrinking of the core may still drive the total rate of annihilations to increase; if so, there is the possibility of igniting a SN at sizes $r \lesssim R_{\chi \chi}$.
We have estimated that this is not the case.
However, as described in Sec.~\ref{sec:AnnihilatingDM}, the evolution of the annihilating DM core here is somewhat complicated and requires more detailed study---thus we only consider the constraints on annihilations while the DM core is still at sizes $r \gtrsim R_{\chi \chi}$.

Of course, the DM core may never annihilate efficiently if it first collapses to a BH $G N_\x{col} m_\chi \gtrsim R_{\chi \chi}$, though the energy deposited by annihilations before the core shrinks to within the Schwarzschild radius may still be sufficient to ignite a SN.
Similarly, if the DM core first reaches the size at which QM effects become important before efficiently annihilating $R_\x{QM} \gtrsim R_{\chi \chi})$, then the energy deposited by annihilations at or before this point may still be sufficient to ignite a SN.
We have included both of these constraints.

We now consider annihilations igniting SN after formation of a fermi degenerate core or a BEC.
As shown in Sec.~\ref{sec:AnnihilatingDM}, a fermi degenerate core shrinks by capturing additional DM and can saturate once the capture rate is of order the annihilation rate.
If this saturation occurs before the core has a chance to shrink much below $R_\x{QM}$, then it does not ignite a SN.
On the other hand if saturation occurs at a number~\eqref{eq:Nchichif} much greater than the initial collapsing number, then annihilations in the fermi degenerate core can ignite a SN at a number $N \lesssim N_{\chi \chi}^\x{f}$.
The energy deposited in a trigger region $\lambda_T^3$ and a diffusion time $\tau_\x{diff}$ is:
\begin{align}
\label{eq:Echichif}
\mathcal{E}_{\chi \chi}^\x{f}(N) &\sim m_\chi \frac{N^2}{r^3} \sigma_{\chi \chi} v_\x{col}(r) \tau_\x{diff} \cdot \x{min} \left[1, \left(\frac{\lambda_T}{r} \right)^3\right], \nonumber \\
r &\sim \frac{1}{G m_\chi^3 N^{1/3}}.
\end{align}
Thus a shrinking fermi degenerate core ignites a SN through annihilations if~\eqref{eq:Echichif} satisfies $\mathcal{E}_{\chi \chi}^\x{f} \gtrsim \Eboom$~\eqref{eq:boom2}.
Of course this assumes that $N \lesssim N_\x{life}$ and that the core has not yet collapsed to a BH first $N \lesssim N_\x{Cha}^\x{f}$.

Similarly, a self-gravitating BEC that is collecting particles from the enveloping non-condensed core will saturate at a number~\eqref{eq:Nchichib}.
This highly compact BEC can ignite a SN at any number $N \lesssim N_{\chi \chi}^\x{b}$.
The energy deposited by annihilations in the BEC within a time $\tau_\x{diff}$ (or~\eqref{eq:tbec}, whichever is shorter) is simply:
\begin{align}
\label{eq:Echichib}
\mathcal{E}_{\chi \chi}^\x{f}(N) &\sim m_\chi \frac{N^2}{r^3} \sigma_{\chi \chi} v_\x{BEC}(r) \tau_\x{diff} \cdot \x{min} \left[1, \left(\frac{\lambda_T}{r} \right)^3\right], \nonumber \\
r &\sim \frac{1}{G m_\chi^3 N}.
\end{align}
and will ignite a SN if it is satisfies $\mathcal{E}_{\chi \chi}^\x{b} \gtrsim \Eboom$~\eqref{eq:boom2}.
Of course this also assumes that the BEC has not yet collapsed to a BH $N \lesssim N_\x{Cha}^\x{b}$.
Note that the DM annihilation cross section must be extremely small for a shrinking BEC to have not ignited a SN before formation of a BH: the requirement $\mathcal{E}_{\chi \chi}^\x{b}(N_\x{Cha}^\x{b}) \gtrsim \Eboom$ implies cross sections as low as $\sigma_{\chi \chi} v_\x{BEC} \gtrsim \frac{\Eboom}{\Mpl^4 \tau_\x{diff}} \sim 10^{-90}~\cm^3/\x{s}$ would ignite a SN through annihilations in the BEC.

\begin{figure}
\centering
\includegraphics[width=8cm]{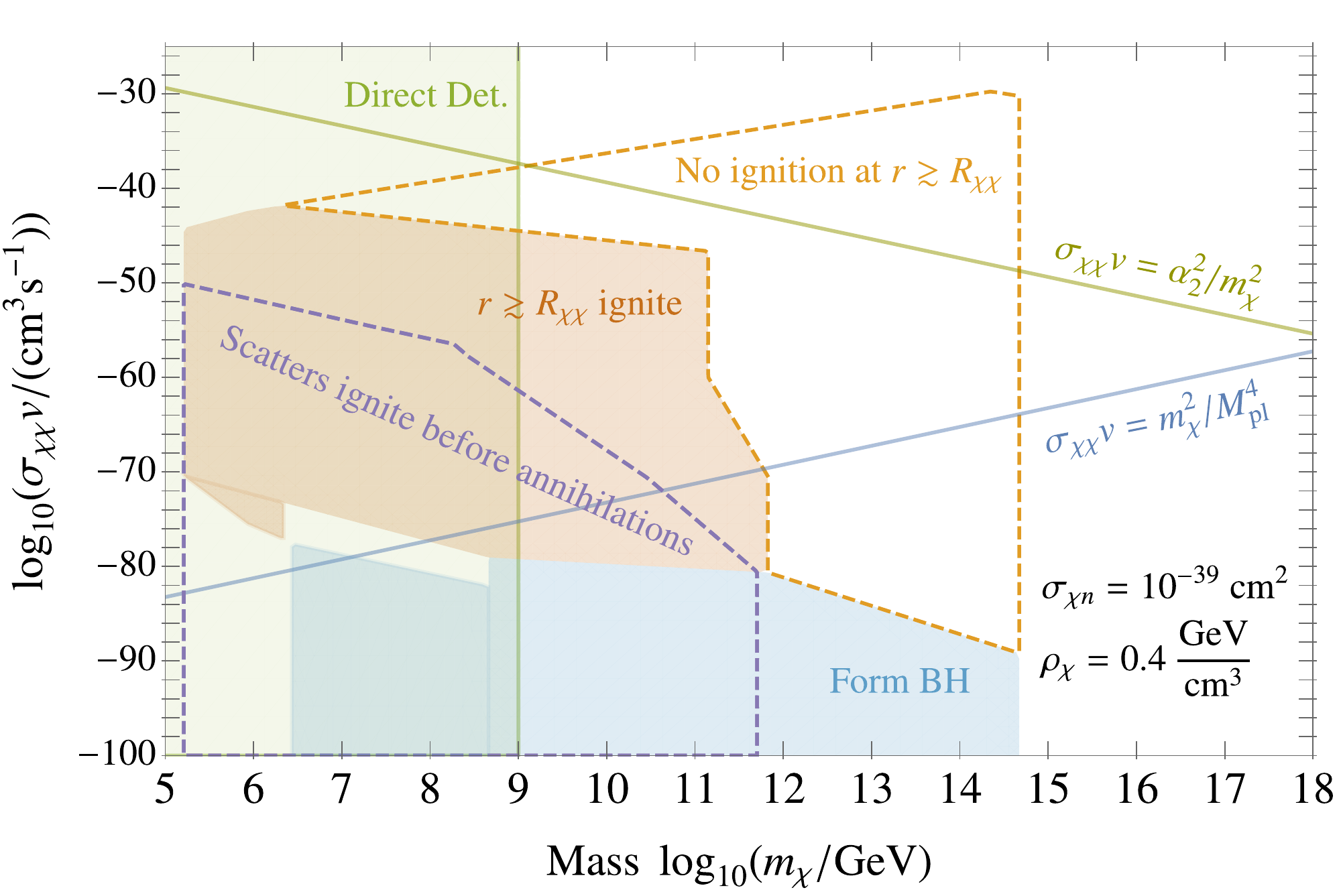}
\caption{Constraints on fermionic asymmetric DM which forms a DM core and ignites a supernova through annihilations ({\color{c6} red}).
For sufficiently small $\sigma_{\chi \chi} v$ the core first collapses to a black hole ({\color{c7} blue}), and is otherwise constrained, see Fig.~\ref{fig:BHfermion}.
Also shown ({\color{c5} purple}) are the constraints on DM-nuclei scatters igniting a supernova during core collapse before annihilations could do so.
}
\label{fig:annfermion}
\end{figure}

\begin{figure}
\centering
\includegraphics[width=8cm]{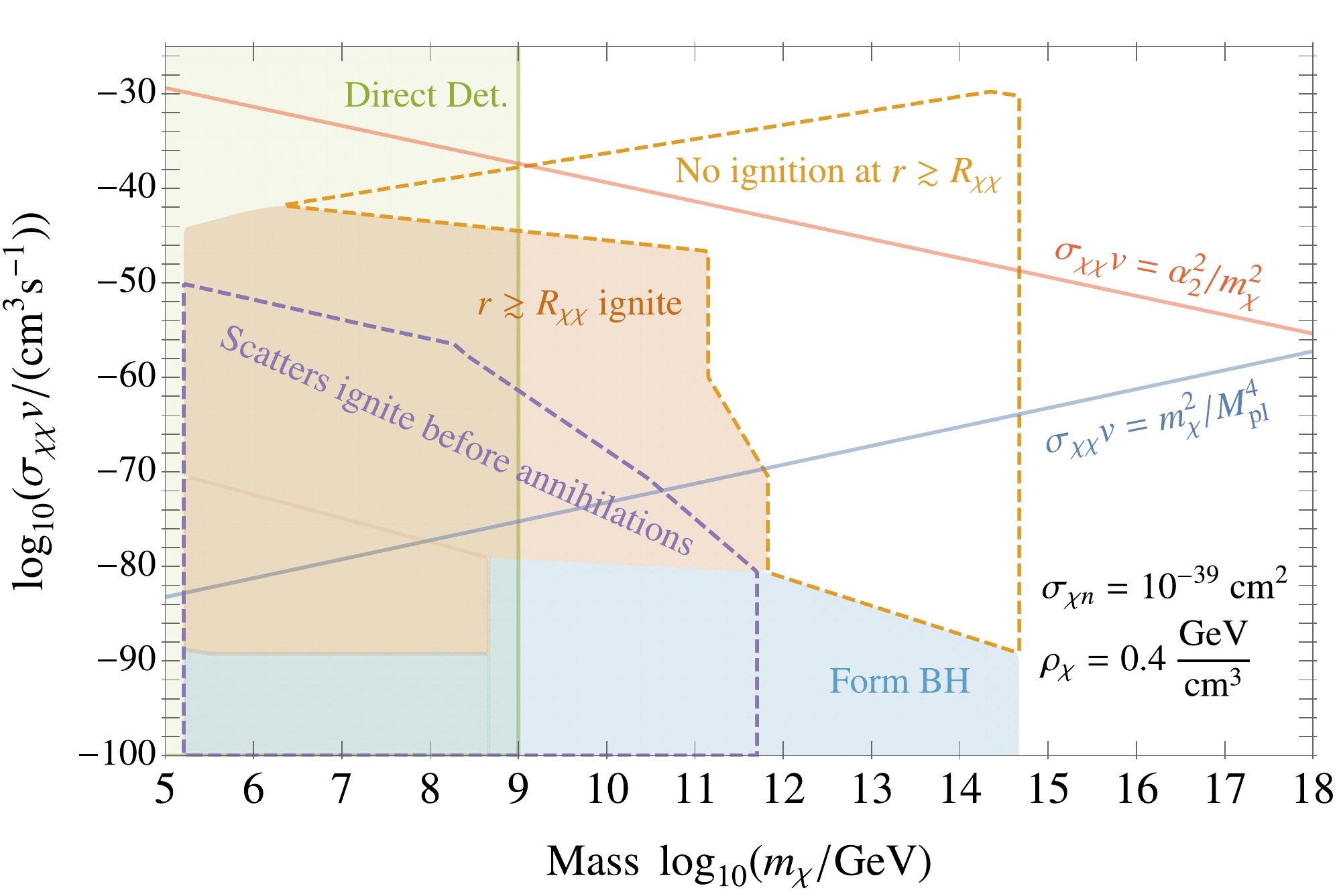}
\caption{Constraints on bosonic asymmetric DM which forms a DM core and ignites a supernova through annihilations ({\color{c6} red}).
For sufficiently small $\sigma_{\chi \chi} v$ the core first collapses to a black hole ({\color{c7} blue}), and is otherwise constrained, see Fig.~\ref{fig:BHboson}.
Also shown ({\color{c5} purple}) are the constraints on DM-nuclei scatters igniting a supernova during core collapse before annihilations could do so.
}
\label{fig:annboson}
\end{figure}

To summarize, a collapsing DM core can ignite a SN by a large number of rapid annihilations.
These constraints are valid regardless of the nature of the annihilation products as long as they deposit their energy within a trigger sized region.
The resulting constraints on DM parameters are shown in Fig.~\ref{fig:annfermion} (fermionic DM) and Fig.~\ref{fig:annboson} (bosonic DM), taking a fixed value of the scattering cross section $\sigma_{\chi n} = 10^{-39}~\cm^2$.
This roughly corresponds to the interaction strength for $Z$-boson exchange, i.e., heavy hypercharged DM (or ``WIMPzilla")~\cite{PhysRevD.64.043503, PhysRevD.59.023501, Feldstein:2013uha, Harigaya:2016vda}.
We also show the constraint from DM-nuclei scatters igniting a SN during core collapse at any point before DM annihilations would have done so.

Note that for an explicit DM model $\sigma_{\chi \chi} v$ is typically related to the DM mass in a calculable way, e.g. s-wave annihilation to electroweak gauge bosons $\sigma_{\chi \chi} v \sim \alpha_2^2/m_\chi^2$ in the case of hypercharged DM.
As shown in Fig.~\ref{fig:annfermion} and Fig.~\ref{fig:annboson}, we constrain annihilation cross sections many orders of magnitude smaller than this naive estimate.
However, this estimate is based upon annihilations of DM its antiparticle $\chi \bar{\chi} \to \x{SM}$, with both existing in roughy equal abundances today.
It is straightforward to imagine a scenario in which essentially no $\bar{\chi}$ particles remain today, and yet $\chi$ is capable of annihilating itself through a parametrically suppressed interaction.
To demonstrate, an explicit DM model of this sort is hypercharged DM with a large vector-like mass and an additional small dimension-5 Majorana mass term (as in the Weinberg operator).
We emphasize though that any DM candidate which can annihilate itself through higher dimension operators may have $\sigma_{\chi \chi} v$ small enough to be constrained by our results e.g., annihilation to SM fermions through a Planck-suppressed cross section $\sigma_{\chi \chi} v \sim m_\chi^2/\Mpl^4$.

\section{Discussion}
\label{sec:discussion}

We have studied the possibility of DM core collapse triggering type Ia SN in sub-Chandrasekhar WDs, following up on previous work~\cite{Graham:2018efk}.
Collapse of asymmetric DM can lead to the formation of a mini BH which ignites a SN by the emission of Hawking radiation, and collapse of annihilating DM can lead to large number of rapid annihilations which also ignite a SN.
Such processes allow us to place novel constraints on DM parameters, as shown in Fig.~\ref{fig:BHfermion}, Fig.~\ref{fig:BHboson}, Fig.~\ref{fig:annfermion}, and Fig.~\ref{fig:annboson}.
These constraints improve on the limits set by terrestrial experiments, and they are complementary to previous considerations of DM capture in compact objects.
It is interesting to contemplate that the ignition of type Ia SN through the evaporation of mini black holes represents a potential observable signature of Hawking radiation.
Further, it also interesting that the extremely tiny annihilation cross sections constrained in this work, which to our knowledge have no other observable consequences, can nonetheless be capable of igniting a SN. 

The processes studied here present a number of opportunities for future work.
The DM constraints presented in this paper are based on the existence known, heavy WDs.
It would also be interesting to calculate the constraints on DM core collapse scenarios arising from the observed galactic SN rate---these may depend more sensitively on the timescale to form a core, or in the case of BH formation, the evaporation time.
In addition, we have restricted our attention here and in~\cite{Graham:2018efk} to DM candidates which interact with the SM through short-range, elastic nuclear scatters.
It would be interesting to broaden our scope to relics with qualitatively different interactions, such as inelastic scatters or radiative processes.
DM which can cool via emission of dark radiation will be more susceptible to collapse, and is likely to be more strongly constrained than models possessing only elastic cooling.
Another particularly interesting case is electrically charged particles~\cite{champs} or magnetic monopoles.
Ultra-heavy monopoles and anti-monopoles could be captured in a WD and subsequently annihilate, igniting SN---we estimate that such a process can be used to place constraints on the flux of galactic monopoles exceeding current limits~\cite{monopoles}.

Finally, though we have not touched upon it here, there are many puzzles in our understanding of the origin of type Ia SN and other WD events, such as Ca-rich transients.
It is plausible (e.g., see the discussion in~\cite{Graham:2018efk}) that DM is responsible for a fraction of these events.
To this end, it is important to identify the distinguishing features of SN that would originate from DM core collapse (e.g. the lack of a stellar companion) in order to observationally test such tantalizing possibilities.

\emph{Note added:}
While this paper was in the final stages of preparation,~\cite{Acevedo:2019gre} appeared which has some overlap with this work.

\section*{Acknowledgements}
We thank Jeff Dror, David Dunsky, Michael Fedderke, Keisuke Harigaya, Chris Kouvaris, Jacob Leedom, Sam McDermott, Surjeet Rajendran, and Petr Tinyakov for useful discussions.

\bibliography{wd}
\end{document}